\documentclass[sn-mathphys,Numbered]{sn-jnl}

\newcommand {\ifHilight} {}

\newcommand {\ifHilightt} {}

\usepackage{graphicx}%
\usepackage{multirow}%
\usepackage{amsmath,amssymb,amsfonts}%
\usepackage{amsthm}%
\usepackage{mathrsfs}%
\usepackage[title]{appendix}%
\usepackage{xcolor}%
\usepackage{textcomp}%
\usepackage{manyfoot}%
\usepackage{booktabs}%
\usepackage{algorithm}%
\usepackage{algorithmicx}%
\usepackage{algpseudocode}%
\usepackage{listings}%
\usepackage{titletoc}%
\usepackage{soul}%
\usepackage{textcomp}%
\usepackage{textgreek}
\usepackage{pifont}
\usepackage{geometry}
\usepackage[resetlabels]{multibib}

\newcites{S}{Supplementary Note References}

\usepackage{lineno}
\usepackage{acronym}
\usepackage{newfloat}
\DeclareFloatingEnvironment[fileext=lop]{Extended Data Fig.}
\DeclareFloatingEnvironment[fileext=lop]{Extended Data Table}

\acrodef{EO}{electro-optic}
\acrodef{HPC}{high-performance computer}
\acrodef{XR}{extended reality}
\acrodef{MPW}{multi-project wafer}
\acrodef{CMOS}{complementary metal-oxide-semiconductor}
\acrodef{CPO}{co-packaged optics}
\acrodef{AI}{artificial intelligence}
\acrodef{ML}{machine learning}
\acrodef{SiP}{silicon photonics}
\acrodef{MZM}{Mach-Zehnder modulator}
\acrodef{MZI}{Mach-Zehnder interferometer}
\acrodef{MRM}{microring modulator}
\acrodef{MRA-MZM}{microring-assisted MZM}
\acrodef{MOS}{metal-oxide-semiconductor}
\acrodef{DUT}{device under test}
\acrodef{TOPS}{thermo-optic phase shifter}

\acrodef{FWHM}{full width at half-maximum}

\acrodef{DAC}{digital-to-analog converter}
\acrodef{DCI}{data-center interconnect}
\acrodef{ADC}{analog-to-digital converter}
\acrodef{AWG}{arbitrary waveform generator}
\acrodef{I/Q}{in-phase/quadrature}
\acrodef{IM-DD}{intensity-modulated direct-detection}
\acrodef{PD}{photodetector}
\acrodef{OMA}{optical modulation amplitude}
\acrodef{ER}{extinction ratio}
\acrodef{FSR}{free spectral range}
\acrodef{TPA}{two-photon absorption}
\acrodef{FCA}{free carrier absorption}
\acrodef{FCD}{free carrier dispersion}
\acrodef{DWDM}{dense wavelength-division multiplexing}
\acrodef{B2B}{back-to-back}
\acrodef{SP}{single polarization}
\acrodef{DP}{dual polarization}
\acrodef{OSNR}{optical signal-to-noise ratio}
\acrodef{ASE}{amplified spontaneous noise}

\acrodef{OBPF}{optical band-pass filter}

\acrodef{VNA}{vector network analyzer}
\acrodef{ECL}{external cavity laser}
\acrodef{EDFA}{Erbium doped fiber amplifier}

\acrodef{PDME}{polarization division multiplexing emulator}
\acrodef{PC}{polarization controller}
\acrodef{VOA}{variable optical attenuator}
\acrodef{SMF}{single mode fiber}
\acrodef{BPD}{balanced photodetector}
\acrodef{RTO}{real-time oscilloscope}
\acrodef{OSA}{optical spectrum analyzer}
\acrodef{SDM}{space division multiplexing}
\acrodef{WDM}{wavelength division multiplexing}
\acrodef{PAM}{pulse-amplitude modulation}
\acrodef{ASK}{amplitude-shift keying}
\acrodef{BPSK}{binary phase-shift keying}
\acrodef{QAM}{quadrature amplitude modulation}
\acrodef{PRBS}{pseudorandom binary sequence}
\acrodef{SPS}{sample-per-symbol}
\acrodef{ISI}{intersymbol interference}
\acrodef{RF}{radio frequency}
\acrodef{DSP}{digital signal processing}
\acrodef{DPD}{digital pre-distortion}
\acrodef{LUT}{look-up table}
\acrodef{RRC}{root-raised cosine}
\acrodef{FIR}{finite impulse response}
\acrodef{LPF}{low pass filter}
\acrodef{CDE}{chromatic dispersion equalizer}
\acrodef{MIMO}{multiple-input multiple-output}
\acrodef{CMA}{constant modulus algorithm}
\acrodef{MMA}{multi-modulus algorithm}
\acrodef{LMS}{least mean square}
\acrodef{CPR}{carrier phase recovery}
\acrodef{FOC}{frequency offset compensation}
\acrodef{BER}{bit error rate}
\acrodef{FEC}{forward error correction}
\acrodef{EVM}{error vector magnitude}
\acrodef{QPSK}{quadrature phase shift keying}
\acrodef{CPO}{co-packaged optics}
\acrodef{SOI}{silicon-on-insulator}
\acrodef{FOM}{figures of merit}
\acrodef{IM}{intensity modulation}
\acrodef{AM}{amplitude modulation}
\acrodef{PM}{phase modulation}
\acrodef{PCB}{printed circuit board}

\begin{document}

\title[Article Title]{Ultrafast Coherent Dynamics of Microring Modulators}


\author[1]{\fnm{Alireza} \sur{Geravand}}\email{alireza.geravand.1@ulaval.ca}

\author[1]{\fnm{Zibo} \sur{Zheng}}\email{zibo.zheng.1@ulaval.ca}
\author[1]{\fnm{Farshid} \sur{Shateri}}\email{farshid.shateri.1@ulaval.ca}
\author[1]{\fnm{Simon} \sur{Levasseur}} \email{simon.levasseur@copl.ulaval.ca}
\author[1]{\fnm{Leslie} \sur{A. Rusch}}\email{leslie.rusch@gel.ulaval.ca}
\author*[1]{\fnm{Wei} \sur{Shi}}\email{wei.shi@gel.ulaval.ca}

\affil[1]{\orgdiv{Department of Electrical and Computer Engineering, Centre d’optique, photonique et laser (COPL)}, 
\orgname{Université~Laval}, \orgaddress{\city{Quebec City}, \state{Quebec}, \country{Canada}}}




\abstract{Next-generation computing clusters require ultra-high-bandwidth optical interconnects to support large-scale artificial-intelligence applications. These electronic-photonic co-integrated systems necessitate densely integrated high-speed electro-optical converters. In this context, microring modulators (MRMs) emerge as a promising solution, prized for their exceptional compactness and energy efficiency. Nevertheless, their potential is curtailed by inherent challenges, such as pronounced frequency chirp and dynamic non-linearity. Moreover, a comprehensive understanding of their coherent dynamics is still lacking, which further constrains their applicability and efficiency.  Consequently, these constraints have confined their use to spectrally inefficient intensity-modulation direct-detection links.
In this work, we present a thorough study of MRM coherent dynamics, unlocking phase as a new dimension for MRM-based high-speed data transmission in advanced modulation formats.
\ifHilightt{We demonstrate that the phase and intensity modulations of MRMs exhibit distinct yet coupled dynamics, limiting their direct application in higher-order modulation formats. This challenge can be addressed by embedding a pair of MRMs within a Mach–Zehnder interferometer in a push-pull configuration, enabling a bistable phase response and unchirped amplitude modulation. Furthermore, we show that its amplitude frequency response exhibits a distinct dependency on frequency detuning compared to phase and intensity modulations of MRMs, without strong peaking near resonance.}
Harnessing the ultra-fast coherent dynamics, we designed and experimentally demonstrated an ultra-compact, ultra-wide-bandwidth in-phase/quadrature (I/Q) modulator on a silicon chip fabricated using a CMOS-compatible photonic process. Achieving a record on-chip shoreline bandwidth density exceeding 5~Tb/s/mm, our device enabled coherent transmission for symbol rates up to 180~Gbaud and a net bit rate surpassing 1~Tb/s over an 80~km span, with modulation energy consumption as low as 10.4~fJ/bit.
}

\maketitle
\newpage
\section*{Introduction}\label{secIntro}

Recent years have witnessed exponential growth in computational demands for \ac{AI} and \ac{ML}, driven by ever more complex algorithms and larger datasets \cite{Rajbhandari2021}. 
Growth governed by Moore's Law \cite{goldstein2023generative, Sevilla2022} was insufficient to provide the required compute capacity in graphics processing unit (GPUs) and central processing unit (CPUs).
Massively parallel processing in computer clusters and data centers \cite{Narayanan2021} was called upon to bridge the gap in capacity.
Disaggregated resources are key  to keep up with complex \ac{AI}/\ac{ML} tasks, including distributed compute nodes and expanded memory. As illustrated in Fig.~\ref{fig:Arch_intro}a, we need a photonic fabric, i.e.,  novel optical interconnect solutions, that is scalable in both bandwidth and distance \cite{Cheng2018,Khani2021}.

By carrying information in both the amplitude and phase of light, as well as orthogonal polarization states, coherent optics has enabled extreme transmission capacity in modern communication systems.
Coherent detection has been widely deployed for inter-data center links, and holds great promise for campus-scale data-center networks \cite{Shi2020, kobayashi2022coherent, zhou2023state}.
For short reach interconnects, low-cost, low-complexity \ac{IM-DD} remains dominant within the data center. However, there is increasing interest in coherent optics for intra-data-center interconnects as they substantially enhance receiver sensitivity. This allows a remarkable improvement in link budget margin, crucial for next-generation optical transceivers targeting 1.6~Tbps and 200~Gbps per lane  and beyond \cite{Hirokawa2021,Zhou2020}. Coherent detection can also enable networks using optical switching for large-scale ML systems and data centers \cite{poutievski2022jupiter}. 

Traditional long-haul coherent transmission systems use complex, costly \ac{DSP}. Recent progress has shown that low-power and low-latency coherent optical links can be realized by offloading the conventional \ac{DSP} into the optical domain, such as carrier recovery, polarization recovery, and bandwidth equalization \cite{Maharry2023}. This power-efficient approach to coherent detection can be leverage for fast optical switches \cite{Saleh2021} via a re-configurable interconnect network such as that in Fig.~\ref{fig:Arch_intro}b.
Other advantages of coherent optics over \ac{IM-DD} include  greater spectral efficiency and tolerance to optical impairments, which together translate to better scaling for larger switch fabrics. 

Coherent optics are poised to appear in the most sophisticated microelectronic chips, such as  application-specific integrated circuits (ASICs) for data center switches \cite{minkenberg2021co} and tensor processing units (TPUs). 
Novel technologies such as \ac{CPO} integrate heterogeneous optics (i.e., optical transceivers) and silicon (i.e., digital electronics) on a single packaged substrate to substantially increase the bandwidth and energy efficiency of high-speed signal inputs/output interfaces \cite{Atabaki2018,Mahajan2022,margalit2021perspective}. 
For these co-packaged electronic-photonic systems, bandwidth density (measured in bits per second per unit length) and energy efficiency (measured in bits per unit energy consumption) are the among most critical \acp{FOM}.

In these \acp{FOM}, integrated coherent optical transceivers lag far behind  their \ac{IM-DD} counterparts \cite{Hsu2024,Fathololoumi2020}. This is mainly due to the challenges associated with achieving compactness in coherent modulators. Almost all the high-speed  \ac{I/Q} coherent transmitters demonstrated to date use \acp{MZM} \cite{Shi2020, kobayashi2022coherent}. These modulators have low chirp, and relatively high bandwidth and linearity. However, at several millimeters of length, they also have a large footprint \cite{Reed2010}. The traveling-wave electrodes often used with MZMs \cite{Li2023}  are sensitive to RF crosstalk \cite{Jiang2018},  severely limiting the minimum distance between adjacent devices. This, combined with large footprint, severely hobbles bandwidth density.

Silicon \acp{MRM} are known for their remarkable compactness and ultra-low energy consumption \cite{Zhang2023,Tossoun2024} thanks to their resonance-enhanced modulation efficiency \cite{xu2005micrometre, Dube-Demers2016}.
They have been widely studied for \ac{IM-DD} short-reach interconnects  \cite{Hu2023, Chan2023, sun2019} and are regarded as critical building blocks for high-bandwidth-density \ac{WDM} photonic links \cite{Xu2006,Yuan2024}.

However, their strong frequency chirp, low \ac{ER}, and the lack of in-depth understanding of the dynamics of \ac{I/Q} \acp{MRM} have discouraged research into coherent transmission  \cite{Chang2017}. \ifHilight{Only a few MRM-based modulators for coherent optics have been demonstrated to date for relatively low data rates (up to 28 Gbaud) \cite{Dong2013, DeValicourt2018, Jo2024} (see Extended Data Table~\ref{tab:comparison} for detailed comparison).}

We provide a thorough study on ultrafast coherent dynamics of \acp{MRM}, unlocking phase as a new dimension for \ac{MRM}-based high-speed data transmission in advanced modulation formats. 

\ifHilightt{We show that the \ac{PM} of an \ac{MRM} exhibits a distinct frequency response than its \ac{IM}. Embedding a pair of \acp{MRM} within a Mach–Zehnder interferometer in a push-pull configuration can effectively overcome the well-known challenge of frequency chirp. We demonstrate this approach simplifies the \ac{AM}, aligning it with the real part of the complex amplitude of the \ac{MRM}, making it well suited for higher-order advanced modulation formats.}

Harnessing the ultra-fast coherent dynamics of this \ac{MRA-MZM}, we designed and experimentally demonstrated an ultra-compact all-silicon \ac{I/Q} modulator with a \ifHilightt{6-dB bandwidth greater than 60~GHz}. 
Our transmission experiment achieved a net rate greater than 1~Tb/s with dual-polarization coherent transmission for up to 80~km. With a footprint on the order of 100~$\mu$m in width, our device shows a record on-chip shoreline bandwidth density greater than 5~Tb/s/mm. 
Because of the wavelength-selectivity of the \ac{MRM}, a chip-scale integrated \ac{WDM} transceiver system can be realized by simply cascading the proposed coherent modulators, as illustrated in Fig.~{\ref{fig:Arch_intro}}c. 
Our results summarized in Fig.~{\ref{fig:Arch_intro}}d and Extended Data Table~\ref{tab:comparison} are among the best demonstrated to date.

\begin{figure}[t]%
\centering
\includegraphics[width=0.95\textwidth]{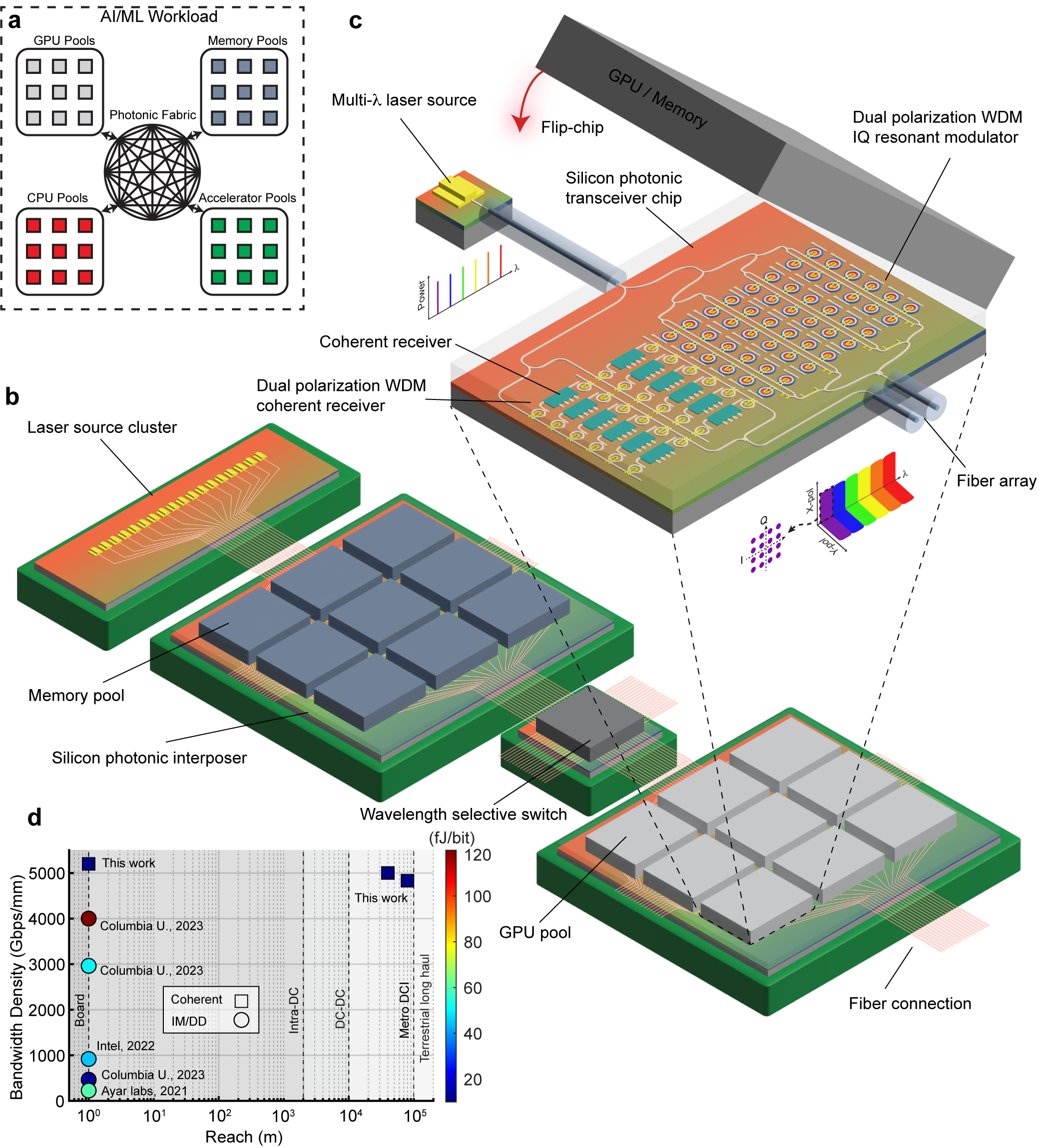}
\caption{\textbf{Conceptualization of ultra-compact coherent silicon photonic links enabling a future disaggregated data center} \textbf{a,} A disaggregated data center scales AI/ML workloads by joining resources via high-performance coherent photonic links. \textbf{b,} Resources  packaged on silicon photonic interposers are connected through optical fibers and wavelength selective switches, exploiting the reconfigurability and scalability of optical links. \textbf{c,} Co-integration of electrical and photonic integrated circuits (PIC) exploits resonator-based, highly-compact transceivers with wavelength and polarization multiplexing; a multi-\textit{$\lambda$} laser source can be shared by the transmitter and receiver through an on-chip power splitter and scale  modulation of many channels. 
Electrical integrated circuits  including GPUs, CPUs, high-bandwidth memory (HBM), or other accelerators can be interconnected through silicon photonic links. 
\textbf{d,} Performance comparison of our device and other state-of-the-art highly compact transmitters \cite{Fathololoumi2022,Sun2020,Daudlin2023,Rizzo2023,Wang2023} in terms of shoreline bandwidth density, reach, and energy consumption per bit.
}\label{fig:Arch_intro}
\end{figure}

\section*{Results}
\subsection*{Complex-amplitude modulation with MRM and MRA-MZM}

Coherent optics involves the modulation and detection of the complex amplitude of light, encompassing both its amplitude and phase. We first delve into the complex-amplitude modulation with solitary MRMs, focusing on the add-drop configuration in single-mode operation, as depicted in Fig.~\ref{fig:mrm_intro}a, where the through-port of the bus waveguide outputs the modulated signal while the drop-port is used for monitoring and stabilization of resonance. 
The coupling status of the bus waveguide to the resonator, in comparison to the resonator round-trip loss, is classified into three specific conditions: under-coupled, critically-coupled, and over-coupled.
We shall see that the phase response of an MRM strongly depends on the coupling condition.

Phasor diagrams of the normalized electric field transfer function are intuitive representations of both the amplitude and phase of a resonator. Figure~\ref{fig:mrm_intro}b presents the phasor diagram of the MRM through-port output for three bus-resonator coupling coefficients resulting in under-coupled, critically-coupled, and over-coupled conditions. The amplitude response measures the distance from the origin, while the phase is represented as the angle determined by the wavelength offset defined by $\Delta\lambda = \lambda - \lambda_r$, where $\lambda$  and $\lambda_r$ are wavelength and resonant wavelength, respectively. 
Note that for small detunings, the phasor in each case passes through resonance (dashed line) rapidly, and the output amplitude reaches zero at resonance only in critical coupling. In under-coupled and critically-coupled cases, the maximum range of achievable phase variation is below and equal to \textpi, respectively. On the other hand, in an over-coupled MRM, the output signal undergoes a 2\textpi\ phase shift across the resonance wavelength, forming a circle around the origin. 
At the same time, the optical signal undergoes amplitude filtering at the resonance wavelength, which is clearly observable as the point on the circle nearest to the origin.
The extent of the amplitude filtering and the gradient of the phase response are dependent on the level of over-coupling.
As the resonator moves away from critical coupling towards more pronounced over-coupling, there is a noticeable reduction in amplitude filtering, accompanied by a decrease in the steepness of the phase response. This change is illustrated by the broadening of the circle and its closer alignment with the unit circle on the complex plane. \ifHilightt{Further details on complex plane representation of \ac{MRM} response for different coupling regimes are given in Supplementary Note~\ref{SN:sec_MRM_coupling}.} 
The capability to encompass all quadrants of the complex plane is a beneficial attribute, rendering over-coupled MRMs a preferred option for various applications \cite{Liang2021}. Therefore, \ifHilight{our attention for the remainder of this study pivots towards the over-coupled condition, where substantial phase and amplitude modulation can be achieved. In all cases studied in the remainder of this work, we consider devices with identical MRM designs, each with a Q-factor of 3,200 for both the solitary MRM and the MRA-MZM. Note that the device parameters are chosen to represent the fabricated devices under test in the following sections.}

Figure~\ref{fig:mrm_intro}~d and f illustrate the variation in the phasor response of \ac{MRM} for three different applied voltages: $V_b - V_p$, $V_b$, and $V_b + V_p$, where $V_b$ and $V_p$ are bias and peak swing voltages, respectively. All under the premise that the p-n junction within the MRM is reverse biased.
The MRM being analyzed shares identical parameters with the one utilized in the prototype, which will be elaborated on in the following section. 
Applying a spectrum of reverse bias voltages modulates the intracavity loss and effective index, therefore causing a subtle shift of the phasor transfer function, as shown in Fig.~\ref{fig:mrm_intro}d and further detailed in Fig.~\ref{fig:mrm_intro}f. While the reduced loss from applying a higher voltage ($V_{in}: V_b - V_p \rightarrow V_b + V_p$) manifests as an increase in the circle radius, the modulated effective index is presented as a rotation of the circle around its center. 
\ifHilightt{This effectively shifts the resonance wavelength of the MRM, which is indicated with a circle in Fig.~\ref{fig:mrm_intro}~d and f.}

\begin{figure}[]%
\centering
\includegraphics[width=0.99\textwidth]{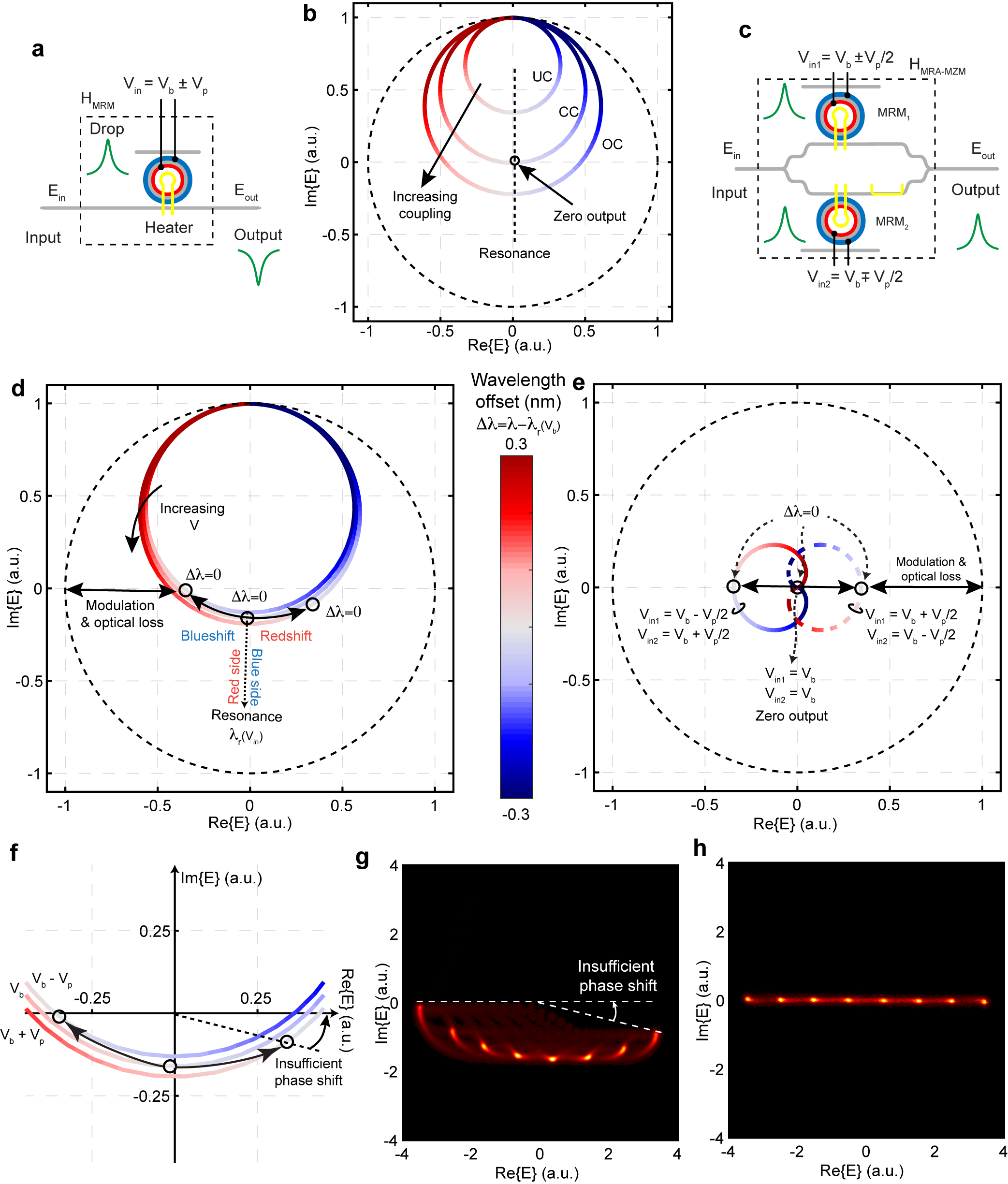}
\caption{\textbf{Operating principle of the microring modulators (MRMs).} \textbf{a,c,} Schematic diagram of an MRM \textbf{(a)} and a microring assisted MZM (MRA-MZM) \ifHilightt{operating at null}\textbf{(c)} driven in push-pull mode. \textbf{b,} Normalized static electric field spectra of MRM on the complex plane for three ring-bus coupling strengths creating under-coupled (UC), critically-coupled (CC), and over-coupled (OC) conditions. \textbf{d,e,f,} Normalized spectra of MRM \textbf{(d,f)} and MRA-MZM \textbf{(e)} on the complex plane for different voltages. 
\ifHilightt{Color bar shows the wavelength offset where \textit{$\Delta \lambda=0$} means the resonance wavelength at bias voltage.} Arrows indicate the trajectory of the modulated signal and the achievable amplitude and \ac{PM}. \textbf{g,h,} Oversampled constellation diagram of a 40~Gbps modulated 8-ASK signal operating at \textit{$\Delta\lambda=0$}, biased at \textit{$V_b$} and a \ifHilight{peak-to-peak swing equal to \textit{$2V_p$}} using MRM \textbf{(g)} and MRA-MZM \textbf{(h)}.
}\label{fig:mrm_intro}
\end{figure}

The strong interplay between phase and amplitude in the MRM results in a significant frequency chirp, manifesting as a curved trajectory of the electric field transfer function on the phasor plane.

This curved trajectory is shown with a curved solid arrow in Fig.~\ref{fig:mrm_intro}d and f, indicating the existence of frequency chirp. In this case, the MRM is biased at $V_b$ with a signal swing of $V_{pp} = 2V_p$ while operating at resonance wavelength (zero laser-resonance detuning, $\lambda_l= \lambda_r(V_b)$). The insufficient voltage swing causes the below full-\textpi\ modulation of the signal. Interestingly, the slope of the transmission spectra is opposite on two sides of the resonance wavelength, leading to opposite chirp signs depending on the detuning sign, which will be useful depending on the application. The finite modulator DC \ac{ER} of the \acp{MRM} also known as resonance depth is also evident since the trajectory does not intersect the origin of the graph, which indicates that the output electric field never reaches zero.

The \ac{MRA-MZM} is composed of a balanced \ac{MZI} with a pair of identical \acp{MRM}, each coupled to one \ac{MZI} arm, plus a \ac{TOPS} that establishes the phase difference between the two arms, as depicted in Fig.~\ref{fig:mrm_intro}c. 
\ifHilightt{Assuming a differential driving signal and a $\phi_b$ phase shift introduced by the TOPS, the complex amplitude of the output electric field can be found to be}: 
\begin{equation}
E_{out}(t) = \frac{1}{\sqrt{2}} (E_\text{MRM1} + E_\text{MRM2}\cdot e^{j\phi_b} ) = \sqrt{2}\underbrace{A_m(t) \ cos \left( \phi_m(t)+\frac{\phi_b}{2} \right)}_{\operatorname{Re}\{-i\cdot E_\text{MRM2}\}} \cdot \ e^{j\frac{\phi_b}{2}}
\label{eq:MZM_out_simp}
\end{equation}
\noindent 
\ifHilightt{where $E_{\text{MRM1}} = A_m(t) \cdot e^{-j\phi_m(t)}$ and $E_{\text{MRM2}} = A_m(t) \cdot e^{j\phi_m(t)}$, with $A_m(t)$, $\phi_m(t)$ being the amplitude and the phase of the modulated optical signal by MRMs, respectively. Assuming operation at null point $\phi_b$ is set to $\pi$.} In Eq.~\ref{eq:MZM_out_simp} the output phase is not a temporal function of input signal.
Therefore, in contrast to the case of solitary MRM, a \ac{MRA-MZM} can have an \ifHilight{infinite \ac{ER}} and zero frequency chirp. This is evident by the straight displacement line in Fig.~\ref{fig:mrm_intro}e, as the differential electrical signal shifts the responses of the \acp{MRM} in opposite directions, minimizing the residual \ac{PM}. The transmission remains constant for $V_b \pm V_p$ but with a $\pi$ phase difference as the driving voltage switches its direction. This explains the substantially suppressed frequency chirp \cite{Chang2017} and nonlinearity \cite{Shawon2023} observed in the experiment, which eventually enables high-speed \ac{ASK} modulation \cite{Geravand2023SiPh}.  The reduction of the modulation chirp can be visually confirmed by comparing the time-domain simulated high-speed modulated signal trajectory in the complex plane for two cases. Here, we have simulated both configurations with an 8-\ac{PAM} signal; the output is plotted on the complex plane in Fig.~\ref{fig:mrm_intro}g and h, confirming the near chirp-free operation of the \ac{MRA-MZM} configuration.

\subsection*{Frequency responses of MRM and MRA-MZM} 
High-speed optical modulators are typically evaluated based on \ac{OMA} and electro-optic bandwidth. In conventional designs, MRMs for \ac{IM} are operated at a frequency detuning where the intensity transmission demonstrates the most pronounced slope, a feature graphically illustrated in Fig.~\ref{fig:mrm_dynamics}a.  
\ifHilightt{Conversely, \ac{PM} is more efficient near the resonance and diminishes as the detuning widens, which can be explained by examining the slope of phase spectral response of \ac{MRM}. The \ac{AM} also achieves its maximum OMA at resonance, highlighting the utilization of the phase response of the optical field.}

\ifHilightt{Our study extends to the simulation of small-signal \ac{EO} responses of MRM and MRA-MZM. Figure~\ref{fig:mrm_dynamics}c presents the simulated 3-dB bandwidths for \ac{IM}, \ac{PM}, and \ac{AM} as functions of the frequency detuning.  
Sensitive to the frequency detuning, \ac{IM} may exhibit strong peaking, resulting in abrupt variations in the \ac{EO} bandwidth near resonance, with more details shown in Extended Data Fig~\ref{fig:mrm_SSEO_num}. 
In contrast, the \ac{AM} does not exhibit peaking and shows a much lower dependency on frequency detuning near resonance. This behavior has also been confirmed in our subsequent experiments.}

\ifHilightt{The \ac{AM} shows near identical OMA and bandwidth for \ac{MRM} and \ac{MRA-MZM} within a wide range of frequency detunings, as expected based on Eq.~\ref{eq:MZM_out_simp}.}
One can use the following equation derived for a two-pole system to estimate the EO bandwidth of the MRA-MZM: 

\begin{equation}
\left( \frac{1}{f_{EO}} \right)^2 = \left( \frac{1}{f_{E}} \right)^2 + \left( \frac{1}{f_{O}} \right)^2 = \left(2\pi R_sC_j\right)^2 + \left(2\pi\tau\right)^2,
\label{eq:EOBW}
\end{equation}

\noindent \ifHilightt{where $C_j$, $R_s$, and $\tau$ are junction capacitance, series resistance, and optical field lifetime \cite{Gheorma2002} of the resonator}. Note that, although widely used, Eq.~\ref{eq:EOBW} does not provide an accurate estimation of EO bandwidth of a solitary MRM, and is often misleading due to the oversimplification of the MRM to a two-pole system without considering the impact of detuning, as illustrated in Fig.~\ref{fig:mrm_dynamics}c. Further details on MRM modeling are provided in Supplementary Note~\ref{SN:sec_MRM_model2}. \ifHilightt{The complete results and details of the numerical study are provided in Methods and Extended Data Fig.~\ref{fig:mrm_SSEO_num}}.

\ifHilightt{To understand their performance for coherent optics, we operate both modulator configurations at zero detuning (See Fig.~\ref{fig:mrm_dynamics}a) and observe their time domain responses with a \ac{BPSK} signal, which represents a large-signal condition. Complex plane representation of the modulated signal in Fig.~\ref{fig:mrm_dynamics}d reveals the zero crossing point displacement in MRM and curved transition of the electric field requiring both real and imaginary parts. The near identical similarity of transitions in time domain amplitude response of \ac{MRM} and \ac{MRA-MZM} in Fig.~\ref{fig:mrm_dynamics}e confirms the observations from Fig.~\ref{fig:mrm_dynamics}b and c. As shown in Fig.~\ref{fig:mrm_dynamics}f and g, the MRA-MZM evidently demonstrates an enhanced \ac{ER} and a superior, near instantaneous phase response.} The frequency modulation chirp is then calculated using the instantaneous phase of the signal. As shown in Fig.~\ref{fig:mrm_dynamics}h, the MRM shows a significant divergence from the carrier frequency over a considerable period of time, while the MRA-MZM shows a discontinuity in chirp at modulation edges due to a sharp transition in phase response.

\ifHilightt{Further details of the time-domain comparison of the two structures in large-signal domain, including the effect of the frequency detuning, are provided in Extended Data Fig.~\ref{fig:mrm_operate} where simulation is conducted across a spectrum of detunings.} 
More information on MRM modeling is also provided in Supplementary Note~\ref{SN:sec_MRM_model2}.

\begin{figure}[]%
\centering
\includegraphics[width=0.99\textwidth]{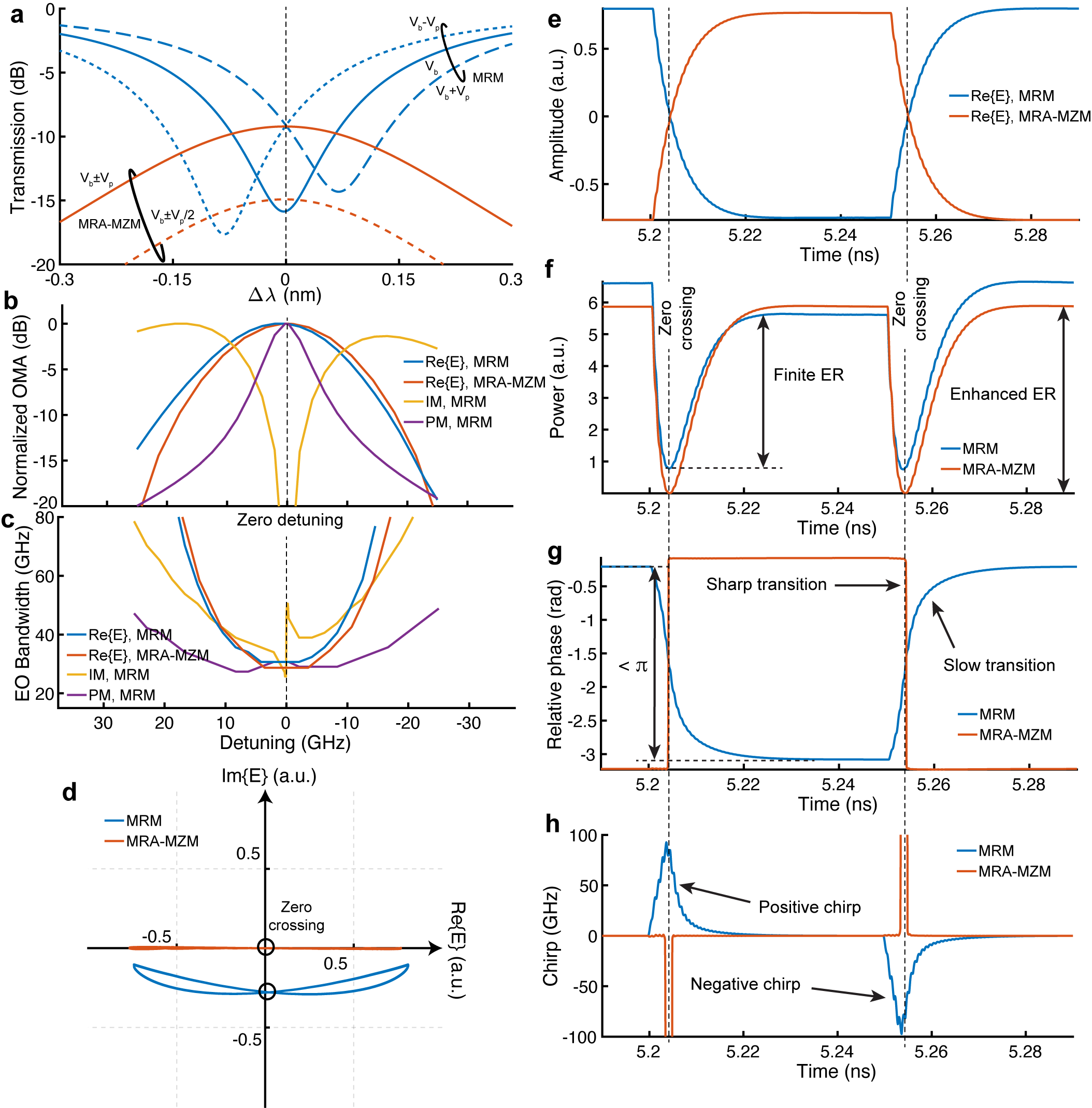}
\caption{\ifHilightt{\textbf{Coherent dynamics of the microring modulators (MRMs).}  \textbf{a,} Transmission spectra of MRM and MRA-MZM. \textbf{b,c,} For MRM and MRA-MZM \textbf{(b)} is normalized small-signal optical modulation amplitude, and \textbf{(c)} is small-signal linear electro-optic bandwidth. Zero detuning is indicated with a dashed line. 
\textbf{d,e,f,g,h,} Large-signal trajectory on the complex plane, amplitude (Real(E)), power, relative phase, and chirp of the modulated BPSK signal for MRM and MRA-MZM for zero detuning indicated in \textbf{(a)}. Displacement of the zero crossing on the complex plane is indicated with circles in \textbf{(d)}.}
}\label{fig:mrm_dynamics}
\end{figure}

\subsection*{Experimental demonstration of chirp-free I/Q modulator}\label{subsec2}

Harnessing the enhanced coherent dynamics of the \ac{MRA-MZM}, we designed an ultra-compact \ac{I/Q} modulator for coherent optical links. 
The device was fabricated using a \ac{CMOS} compatible silicon photonics foundry process (Fig.~\ref{fig:prototype_intro}). This modulator consists of two \acp{MRA-MZM} nested in an \ac{MZI} structure (see Methods). Each \ac{MRA-MZM} is \ac{ASK} modulated to jointly generate  a \ac{QAM} signal (Fig.~\ref{fig:prototype_intro}a). The area occupied by the active components on the chip, including the \acp{MRM}, photodetectors, and \acs{TOPS}, has merely 100~\textmu m in width. This dimension does not account for the routing waveguides (Fig.~\ref{fig:prototype_intro}b-c). While the current demonstration covers only one wavelength channel, all essential functionalities required for \ac{WDM} transmission, such as resonance tuning and monitoring, are integrated. Thanks to their ultra-compact footprints, \ac{DWDM} can be realized by cascading multiple \acs{MRM} along the \ac{MZI} arms, with each modulator column selectively targeting a specific wavelength, as illustrated in Fig.~\ref{fig:Arch_intro}c.

\begin{figure}[]%
\centering
\includegraphics[width=0.99\textwidth]{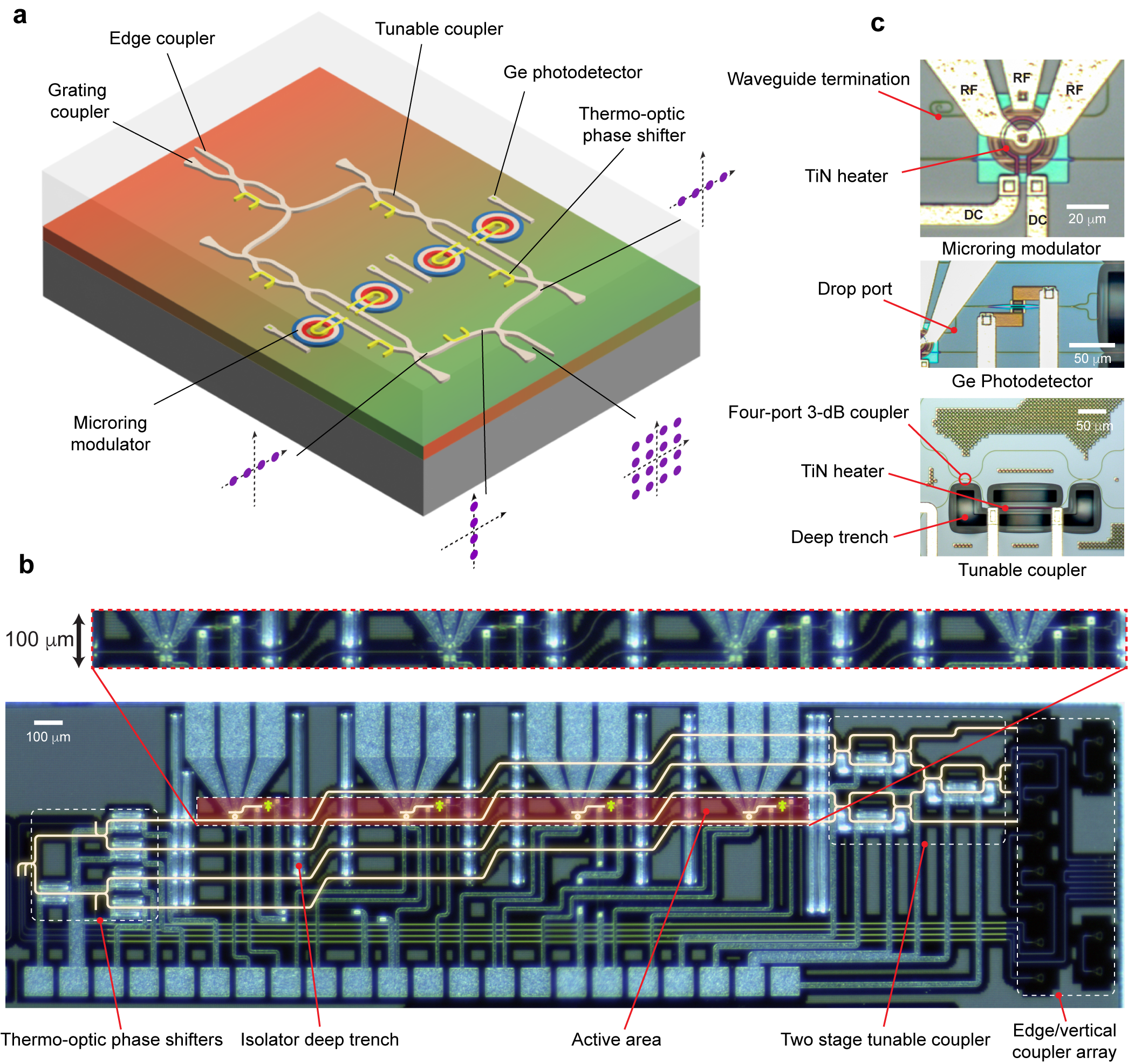}
\caption{\textbf{Overview of prototype for the scalable system architecture and fabricated coherent transmitter chip.} \textbf{a,} 3D visualization of the prototype schematic used in this work with insets of the constellation diagrams of each branch to show the operation principle. \textbf{b,} Micrograph of the die before packaging with annotations and highlights indicating different parts of the circuit, including the active area with a width of 100~\textmu m. \textbf{c,} Microscope images of the microring modulator, Germanium photodetector, and tunable coupler with annotations of their building blocks.
}\label{fig:prototype_intro}
\end{figure}

\subsubsection*{Direct-current (DC) and Electro-optic (EO) characteristics}\label{DCEOsubsection}

To evaluate and assess the performance of the proposed design, initial measurements were conducted on an individually placed \ac{MRM} that shares identical parameters with the ones used in the \ac{MRA-MZM}. \ifHilight{The transmission and phase responses of the modulator under low power conditions (below $-10$~dBm) at various reverse bias voltages are presented in Fig.~\ref{fig:exp_res}a,b along with the simulation results of the modeled device (dashed line). The $V_\pi$ of the MRM as a phase shifter is measured to be around 6.28~V near the resonance, as shown in Extended Data Fig.~\ref{fig:exp_res_2}a}. Note that the same model was used in the simulations of the previous section. The MRM exhibits a DC \ac{ER} of $\sim$17~dB, a \ac{FWHM} of 60~GHz, and a \ac{FSR} of 9.73~nm. The 2\textpi\ phase change across the resonance indicates overcoupling condition. The loss for the PN-junction-loaded waveguide phase shifter is $\sim$85~dB/cm. The power coupling ratio between the resonator and the bus waveguide is estimated to be $\sim$15\%. A Q-factor of 3,200 is extracted from the DC response of the MRM. The relatively high loss of the PN-junction-loaded waveguide phase shifter is not detrimental, as a short photon lifetime are essential for high-bandwidth modulation according to Eq.~\ref{eq:EOBW}. A modulation efficiency of 0.96~V.cm at 0.5~V reverse bias is measured based on $V_{\pi}L=\frac{FSR\cdot\pi R}{\Delta\lambda/\Delta V}$, resulting in a resonance shift efficiency of $\sim$4~GHz/V. The MRM shows a resonator finesse of 20.2 ($F=\tfrac{FSR}{FWHM}$), which can support seven WDM channels with one FWHM spacing between channels as the guard band or twenty zero-guard-band channels.

\begin{figure}[]%
\centering
\includegraphics[width=0.99\textwidth]{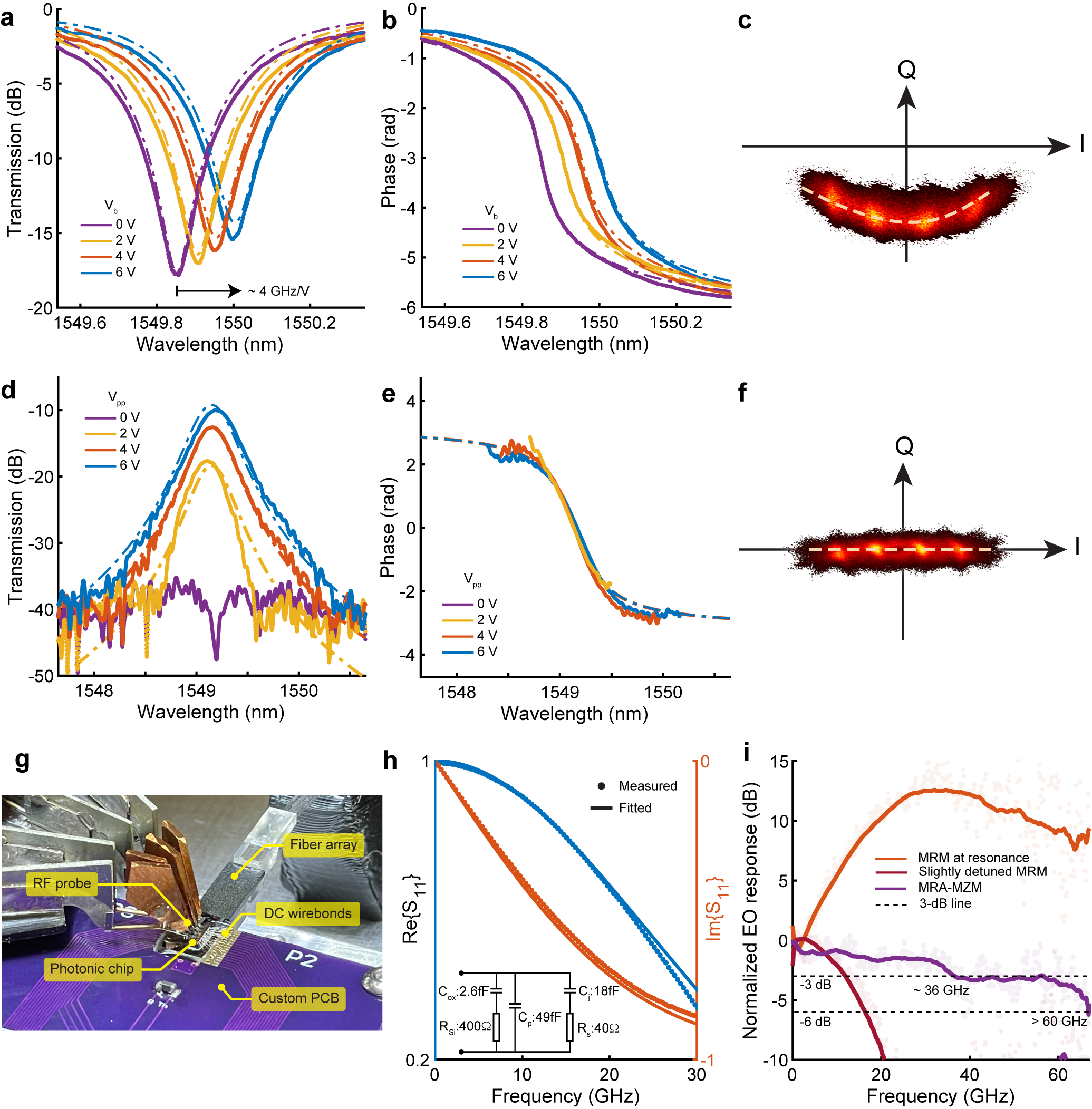}
\caption{\textbf{Experimental characterization results.} \textbf{a,b,} Measured (solid) and simulated (dashed) normalized transmission and phase spectra of the \ac{MRM} with different bias voltages. \textbf{c,} Measured oversampled constellation diagram showing a curved transition trajectory for the \ac{MRM}. \textbf{d,e,} Measured (solid) and simulated (dashed) normalized transmission and phase spectra of the \ac{MRA-MZM} with different bias voltages. \textbf{f,} Measured oversampled constellation diagram showing a straight transition trajectory for the \ac{MRA-MZM}. \textbf{g,} Packaged chip under high-speed RF probing. \textbf{h, }S\textsubscript{11} measurement results and fitted equivalent circuit.\ifHilightt{ \textbf{i, }Measured normalized small-signal linear electro-optic (EO) S\textsubscript{21} responses of the \ac{MRM} at resonance, the slightly detuned \ac{MRM}, and the \ac{MRA-MZM} at zero laser-resonance detuning operating at quadrature point, with dots representing measured data and the solid line representing the smoothed curve. See Supplementary Note~\ref{SN:subsec_linear_SS} for more detail.}
}\label{fig:exp_res}
\end{figure}

 \ifHilight{Figures~\ref{fig:exp_res}d and \ref{fig:exp_res}e present the measured and simulated transmission and phase spectra of the MRA-MZM for a range of driving voltages, showing an extinction ratio greater than 30~dB. We can observe that the spectral response of the phase remains centered at the same wavelength with the varied voltage, confirming its chirp-free operation. The modulation loss of the MRA-MZM is dependent on the swing of the driving signal. This can be seen in Fig.~\ref{fig:mrm_dynamics}a, where the transmission of the MRA-MZM at zero detuning is equal to the transmission of the two oppositely detuned MRMs.}

Each MRM uses an on-chip TiN micro-heater to control its resonance wavelength. The MRM heater exhibits a half-wave power (P\textsubscript{\textpi}) of $\sim$60~mW \ifHilight{(Extended Data Fig.~\ref{fig:exp_res_2}b)}, which can be significantly improved by thermally isolating MRMs from the surrounding area using a silicon undercut process \cite{Coenen2022}. 
Laser-resonance alignment was maintained automatically throughout the experiment by monitoring the drop port with on-chip Ge-on-silicon photodetectors (Fig.~\ref{fig:prototype_intro}c) in a feedback control loop using the perturb-and-observe control algorithm. Details about the control mechanism are provided in Supplementary Note~\ref{SN:sec_stab}.

The relatively high third-order nonlinearity of silicon, coupled with the buildup of optical intensity within the MRM cavity (around 7.3 times in this case), may cause considerable alterations in its spectral responses at high input optical power levels.
Modeling of the MRM in presence of high optical power thus involves the complex interplay between the optical nonlinear and self-heating effects \cite{DeCea2019}. 
The simulated and measured spectral responses, both in amplitude and phase at various optical power levels, are presented in \ifHilight{Extended Data Fig.~\ref{fig:exp_res_2}c and d}. The coupling condition remains over-coupled as the input optical power increases, which is evidenced by the $2\pi$ phase shift across the resonance in the spectrum.
\ifHilight{Notably, under high input optical powers, the phase response exhibits a steeper transition across the resonance. However, this steeper transition is caused by nonlinear phenomena that have a slower time response than the modulation interval, hence it does not necessarily enhance the modulation efficiency in high-speed operation.} Further details on our model are provided in Supplementary Note~\ref{SN:subsec_power_sen}.

The resistance and capacitance of the PN-junction are extracted by fitting an equivalent circuit model to the measured S-parameter ($S_{11}$ in \ifHilight{Fig.~\ref{fig:exp_res}h)}. We estimate the junction series resistance and capacitance to be 40~\textOmega\ and 18~fF, respectively. The RC-limited 3-dB bandwidth of the PN junction is estimated to be 222~GHz. The details of the S-parameter measurements are presented in Supplementary Note~\ref{SN:sec_EO_char}. 

\ifHilightt{The measured normalized small-signal \ac{EO} responses of the \ac{MRM} and \ac{MRA-MZM} are presented in Fig.~\ref{fig:exp_res}~i. The \ac{MRM} exhibits a pronounced peak exceeding 10 dB near 35 GHz, while the \ac{MRA-MZM} demonstrates a smooth roll-off with a 3-dB bandwidth around 35~GHz and a 6-dB bandwidth exceeding 60~GHz. Nonlinear responses, such as second harmonic generation, are not considered in this analysis, as only linear S-parameters can be measured using the linear \ac{VNA}. Further details on the numerical simulations and experiments of the \ac{EO} responses are presented in Supplementary Note~\ref{SN:subsec_linear_SS}.}

The experimental validation of the chirp-free performance of the \ac{MRA-MZM} in comparison to the \ac{MRM} was conducted by applying a differential \ac{QAM} signal and reconstructing the complex optical waveform detected using a coherent receiver \cite{Geravand2023SiPh}. Measurement results, depicted in \ifHilight{Figs.~\ref{fig:exp_res}c and \ref{fig:exp_res}f}, reveal a linear transition trajectory for \ac{MRA-MZM}, in stark contrast to the curved trajectory for the \ac{MRM}. This demonstrates the near-zero chirp characteristic of the \ac{MRA-MZM}. The device tested in our experiments is shown in \ifHilight{Fig.~\ref{fig:exp_res}g} (see Methods for further details).

\subsubsection*{Coherent transmission performance}\label{DataTranssubsection}

Using the proposed chirp-free \ac{MRA-MZM}-based \ac{I/Q} modulator, we performed coherent transmission in advanced modulation formats, including \ac{QPSK}, 16\ac{QAM}, and 32\ac{QAM}. Details of the experimental setup and offline \ac{DSP}  can be found in Methods and Extended Data Fig.~\ref{fig:trans_exp_setup}.

The single-polarization, back-to-back results are presented in Fig.~\ref{fig:exp_trans_res}a, b, and c. In Fig.~\ref{fig:exp_trans_res}a, we sweep symbol rate and report estimated \ac{BER} results. The highest  symbol rate examined was  180~GB, demonstrating the wide bandwidth of the \ac{MRM} under test. At this baud rate the transmitter respected the 24\% soft decision \ac{FEC} threshold for \ac{QPSK}. Similarly, 150~GB 16QAM and 130~GB 32QAM respected the same \ac{FEC} threshold. We translate symbol rate into net bit rate, and present results in Fig.~\ref{fig:exp_trans_res}b. The recorded highest net bit rate is 524.19~Gb/s/pol, given by 130~GB 32QAM with a 24\% \ac{FEC}. To the best of our knowledge, this is the highest single-lane net rate using silicon \acp{MRM}. We have established the potential for over 1~Tb/s with \ac{DP}, which will be verified in a dual-polarization transmission. We select three (symbol rate, modulation order) pairs for the \ac{BER} vs. \ac{OSNR} curves in Fig.~\ref{fig:exp_trans_res}c.  Maximum achievable \ac{OSNR} for 140~GB 16QAM and 120~GB 32QAM are 30.3~dB and 31~dB, respectively, which are found in measurement with lowest \ac{BER}. The \ac{OSNR}-\ac{BER} curves can be interpolated to find the required \ac{OSNR}  to achieve a given \ac{FEC} threshold.

\begin{figure}[]%
\centering
\includegraphics[width=0.99\textwidth]{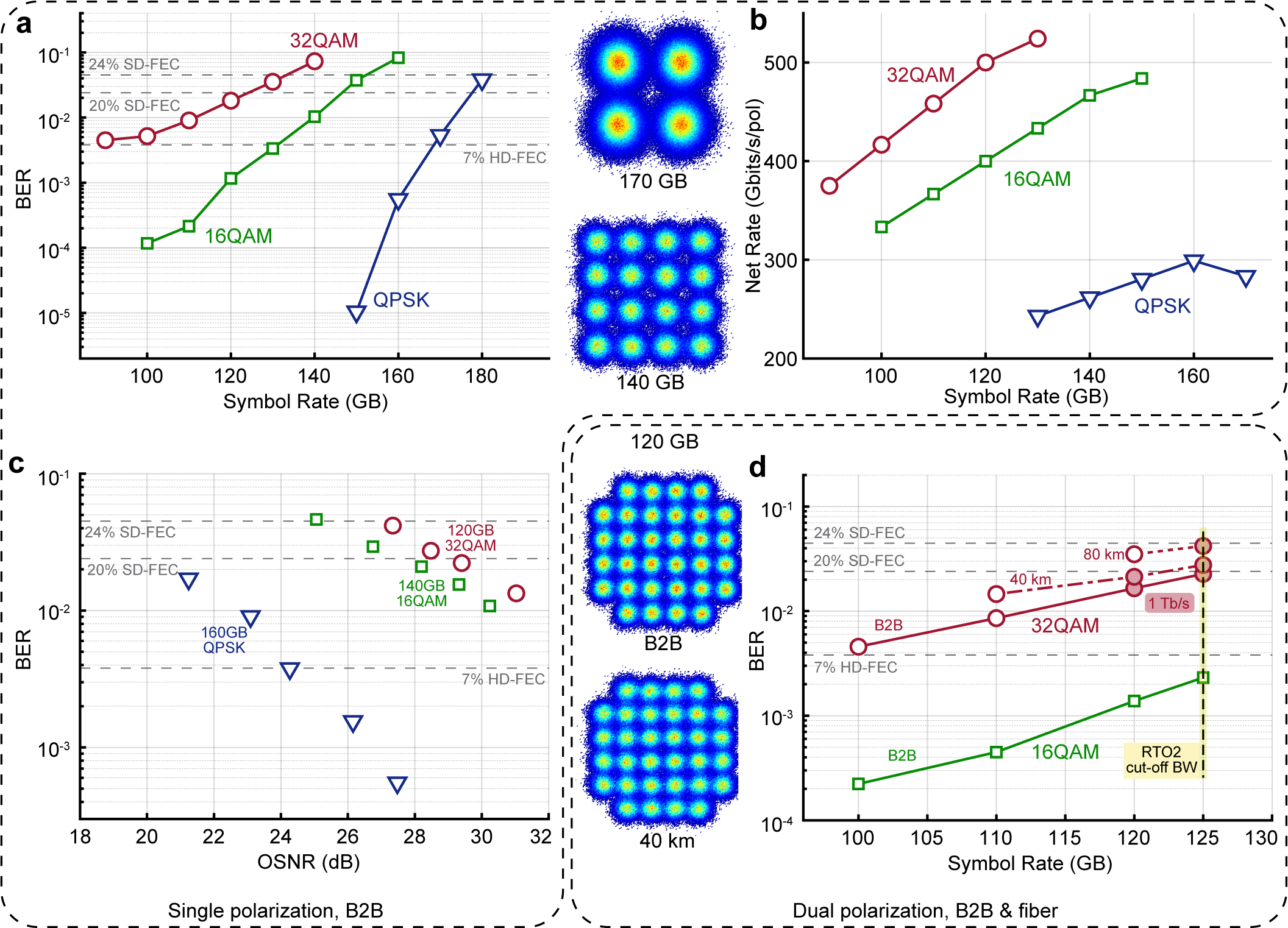}
\caption{\textbf{Coherent data transmission results.} \textbf{a,b,c,} Single polarization (SP) back-to-back \textbf{(a)} BER vs. symbol rate with typical constellations, \textbf{(b),} net bit rate vs. symbol rate,
\textbf{(c)} BER vs. OSNR curves for 160GB-QPSK, 140GB-16QAM and 120GB-32QAM; and  \textbf{d,} dual polarization (DP) results and typical constellations in B2B and over 40~km; shaded markers indicate 1~Tb/s net transmission in B2B and over 80~km.
}\label{fig:exp_trans_res}
\end{figure}

The \ac{DP} results are presented in Fig.~\ref{fig:exp_trans_res}d. The 32QAM BER remains below the 24\% \ac{FEC} threshold up to 125~GB with  B2B, 40~km and 80~km scenarios. We achieve 1~Tb/s or greater on net data rate at shaded markers of 32QAM. For 16QAM, all B2B measurements have \ac{BER} below the 7\% \ac{FEC} threshold. Constellations of 120~Gb 32QAM with B2B and 40~km are also shown. By employing only linear \ac{DSP}, we obtain a symmetric, square constellation with circular symbol clusters. That, again, demonstrates the high linearity of \ac{DUT}. 
The dual-polarization results with back-to-back (B2B) configuration align well with expectations based on the single-polarization measurements illustrated in Fig.~\ref{fig:exp_trans_res}a, suggesting dual-polarization can indeed effectively double the net data rate, as demonstrated in Figure 6c. 

Our transmission results do not reveal the full capability of the \ac{MRA-MZM}. The bandwidth of our test equipment becomes one of the main limiting factors after 140~GB, namely, beyond 70~GHz, rising from uncalibrated RF probes and PDs (see Methods). Noise enhancement by deep digital pre-compensation also is another significant limitation. Besides, power-dependent nonlinearity prevents us from utilizing higher optical laser power to further enhance \ac{OSNR}, which is required from a higher order modulation format. 

Based on the data rate reported in Fig.~\ref{fig:exp_trans_res}d and considering ultra-compact size of the MRMs, our device achieves an exceptional bandwidth shoreline density exceeding 5~Tb/s/mm. This performance is compared with other reported studies in Fig.~{\ref{fig:Arch_intro}}d, where we highlight our achievement of the highest bandwidth density recorded, not only for B2B but also over an 80~km span, indicating that capacity of the MRA-MRM supporting various application scenarios ranging from inter-chip to intra-data-center interconnects. Additionally, using the extracted junction capacitance and the voltage swing employed in the data transmission experiment, we calculated the effective power consumption for modulation to be 10.4 fJ/bit for 32QAM and 18.2 fJ/bit for 16QAM, as detailed in the Methods section. The energy efficiency of our modulator per bit is among the best reported for silicon modulators in the literature.

\section*{Summary and Outlook}\label{secDisc}

We have investigated the coherent dynamics of MRMs and MRA-MZMs, highlighting the potential to exploit phase (the under-exploited dimension of MRM-based devices) for advanced coherent transmission systems. 
Our \ac{MRA-MZM}-based \ac{I/Q} modulator has successfully achieved the first demonstration of Tb/s coherent transmission using a micro-meter-scale, all-silicon modulator. Functioning at the forefront of transmission rate with a remarkable compact footprint, our device has set a new record for bandwidth density per shoreline length, while simultaneously showcasing one of the best modulator energy efficiencies  reported in the literature.

Looking ahead, the potential for scaling up bandwidth density can be significantly enhanced by integrating a wavelength division multiplexing (WDM) system based on the \acp{MRA-MZM}. We proposed a chip-scale coherent \ac{WDM} transceiver, as depicted in Fig.~\ref{fig:Arch_intro}c, which consists of an array of \ac{MRA-MZM}-based \ac{I/Q} modulators alongside an array of coherent receivers, each equipped with an on-chip microring filter. The multi-wavelength laser source could serve both the \ac{I/Q} modulator array and the coherent receiver array through an on-chip power splitter. This source could take the form of either a laser array or an optical frequency comb \cite{Hu2021, Chang2022,Xiang2023}.
Using the proposed design, seven to twenty channels can fit in the FSR (depending on the channel spacing exploited), leading to an aggregated transmission rate of 7-20~Tbps per fiber.
To transcend the limitations imposed by the \ac{FSR} and thus boost the total capacity of the proposed \ac{WDM} architecture, strategies such as leveraging multi-FSR arrangements and interleaved configurations can be employed  \cite{Rizzo2023, Rizzo2021}. The high spectral efficiency afforded by coherent detection allows for a denser packing of channels to maximize the utilization of the optical spectrum.

Leveraging advanced multi-dimensional modulation techniques across a broad spectrum, ultra-high-density coherent optical links can be realized. By using the proposed coherent \ac{WDM} transceivers we facilitate efficient, reconfigurable wavelength selective switches, as envisioned in Fig.~\ref{fig:Arch_intro}b. This advancement not only has the potential to greatly extend the communication reach, it also can support the growth of large-scale disaggregated data centers for AI/ML applications. This approach represents a substantial leap forward in optical communication technology, promising enhanced performance and scalability for future networking infrastructures.

\backmatter

\section*{Methods}

\bmhead{Device design and simulations}

Ansys Lumerical finite difference eigenmode solver is used in designing waveguiding structures. A finite difference time domain solver is used to extract the devices' S-parameters. The extracted parameters are then used to create time-domain compact models in Lumerical INTERCONNECT. Measurement data obtained from DC, RF, and EO characterization are used to match the time-domain circuit model parameters to that of the fabricated device. \ifHilightt{Simulation method to extract the small signal frequency domain characteristics of the devices, including optical modulation amplitude and bandwidth are described in detail in Supplementary Note~\ref{SN:subsec_linear_SS}}

\bmhead{Silicon photonic chip fabrication and design}

The \ac{MRA-MZM} \ac{I/Q} modulator utilizes four identical microrings in an add-drop configuration coupled to two nested \acp{MZI}, which are fed with two-stage tunable couplers, enabling full splitting ratio control to compensate for fabrication errors and to maximize the \ac{ER}. Two micro-heaters control the phase difference and set the operating point of the \acp{MRA-MZM} (see Fig.~\ref{fig:prototype_intro}). A micro-heater controls the phase difference between the I and Q arms. Each MRM has an add-drop racetrack shape with a radius of 10~\textmu m, a coupler length of 600~nm, and a coupler gap of 200~nm. 
The MRMs are designed in the over-coupling regime for a smooth 2\textpi\ phase shift across the resonance. A TiN heater is placed atop the MRM to tune the resonance of each MRM to accommodate fabrication errors.

The designed silicon chip was fabricated using a standard foundry process on a 220~nm-thick \ac{SOI} wafer at Advanced Micro Foundry (AMF) through the MPW service provided by CMC Microsystems. 
The lateral PN junction is formed with six doping layers with doping levels $\sim 2-7\times 10^{18}cm^{-3}$ for the waveguide core to boost the optical bandwidth.
Thermal crosstalk between the rings is minimized by adding on-chip deep trenches: limiting the horizontal heat transfer in the chip.

\bmhead{Device packaging and probing}

The silicon photonic chip was first die-bonded to a custom-designed \ac{PCB}. The DC pads of the photonic chip were wire-bonded to the PCB, while the RF pads were left exposed for RF probing. We exploit edge coupling for optical coupling to and from the chip. A glass interposer is attached to the chip, and the fiber array converts the mode field diameter and pitch of the standard SMF-28 fiber array to inverse taper edge couplers designed on the chip. \ifHilight{The packaged chip (see Fig.~\ref{fig:exp_res}g) exhibits approximately 6~dB of optical passive loss due to fiber-to-chip coupling ($\sim$5~dB) and on-chip waveguide routing ($\sim$1~dB).} 
The transmission and phase response of the device was measured with an optical vector analyzer (LUNA OVA5000).

\bmhead{Transmission setup}

The transmission experiment setup can be found in Extended Data Fig.~\ref{fig:trans_exp_setup}. We demonstrate both \ac{SP} and \ac{DP} with the same transmitter. We explore the highest achievable symbol rate with a \ac{SP}, and we demonstrate terabit capacity with \ac{DP}. 

We use a \ac{DAC} with two differential channels (Keysight M8199B, 80~GHz) to drive two push-pull microring pairs on the chip. We have a phase shifter (Spectrum Elektrotechnik GmbH, 67~GHz) and a bias-T on each RF data path. The phase shifter is used to fine-tune the phase delay between differential channels and bias-T (Anritsu, 65~GHz) provides reverse bias on MRM. RF signals are applied to the chip via a 50~GHz (2.4~mm connector) four-channel GSG probe with 50~\text{$\Omega$} termination.

\ifHilight{The driver in our experiment outputs a swing of 2.7~V per channel before the pre-compensation of the responses of the RF components. The strength of the pre-compensation varies with symbol rate, leading to the voltage swing applied to each MRM ranging from 1 to 2.3~V (See Extended Data Fig.~\ref{fig:exp_res_2}e and f). \ifHilight{Considering a driving swing of 2.2~V$_{pp}$ after the pre-compensation per MRM, the modulation loss is estimated to be $\sim$12~dB at 100~GB, excluding the passive insertion loss of the chip.}}

For the optical transmitter, we use a laser (IDPhotonics CoBrite DX1, line width $<$ 25~kHz) with a center wavelength of 1550.5~nm and boost its power to $\sim$21~dBm by a high-power \ac{EDFA}. A polarization controller and a polarization beam splitter are used to maintain and align the polarization of the beam. 

In the \ac{SP} configuration, we have a 2-stage \ac{EDFA} to amplify the signal power after the silicon chip. \Ac{ASE} noise loading and an \ac{OBPF} (Waveshaper) are placed in between two-stage amplification. The \Ac{ASE} noise loading simulates environments with variable \ac{OSNR}; it consisting of an \ac{EDFA} and a 50:50 coupler. The optical filter is programmed to have a flat response with a 1.8~nm bandwidth centered at the laser wavelength. A 98:2 coupler splits off 2\% power to an optical spectrum analyzer (OSA) to monitor the spectrum. A polarization controller minimizes signal power leakage to another polarization. A variable optical attenuator limits 2~dB maximum power to the optical hybrid. The local laser (IDPhotonics CoBrite DX1) is set to 15~dBm. After coherent detection, we have two \ac{BPD}s (Finisar, 70~GHz) for optoelectronic conversion. We use a two-channel \ac{RTO} (Keysight UXR, 256~GSa/s, 113~GHz BW) to capture the received signal. 

In the \ac{DP} configuration, we use a \ac{PDME} (Kylia) to emulate a \ac{DP} signal. We keep the same two-stage \ac{EDFA} with a \ac{OBPF} at mid-stage. The \ac{VOA} limits \ac{SMF} launch power   to 6~dBm; in the back-to-back case the \ac{VOA} limits the power to the hybrid to 4~dBm. The \acp{BPD} are the same as those for \ac{SP}. We capture the signal with a four-channel \ac{RTO} (Keysight DSOZ634A, 63~GHz) after the coherent detection.

\bmhead{Digital signal processing}

The DSP stacks at TX/RX are also illustrated in Extended Data Fig.~\ref{fig:trans_exp_setup}.
We employ only linear \ac{DSP} techniques at both TX/RX. 

In TX \ac{DSP}, we transmit \ac{PRBS} (PRBS25) pattern and map it to QAM by gray coding. We shape the \ac{QAM} signal with \ac{RRC} pulse of 0.01 roll-off factor. The waveform then is resampled to the \ac{DAC} sampling rate. Waveform correction is applied by Keysight open source software IQ tools. The correction response is depicted in Extended Data Fig.~\ref{fig:trans_exp_setup}, which is estimated from the electrical back-to-back test with phase shifter and bias-T embedded. 

In RX \ac{DSP}, to filter out-band noise we have a $10^{th}$-order low-pass Gaussian filter with a bandwidth of 1.01 times the symbol rate. The \ac{CDE} is applied only when  fiber is present; it is implemented in the frequency domain to invert the dispersion transfer function. Pattern synchronization is done by correlation between the reference signal and captured data. The \ac{MIMO} filter consists of a 4-by-4 real-valued butterfly filter bank found via blind adaptation by the multi-modulus algorithm. The MIMO filters have 133 taps with T/4-spacing. The initial tap coefficients are pre-trained over 8192 symbols before starting the multi-modulus blind equalization. We implement the \ac{FOC} with a fourth power algorithm; we implement a \ac{CPR} via a blind phase search algorithm. A post-\ac{MIMO} filter with seven taps mitigates distortion between \ac{I/Q} coordinates.

\bmhead{Optical spectrum and OSNR measurement}
We monitor the optical signal spectrum via an \ac{OSA} to ensure null-point operation and for \ac{OSNR} measurement. We also  observe bandwidth roll-off from silicon MRM and probes. We present the signal spectrum of 120~GB, 140~GB, and 170~GB in Extended Data Fig.~\ref{fig:trans_exp_setup}.

We estimate \ac{OSNR} from the spectrum,  defined as the ratio of signal power and noise power (noise power within a 0.1~nm resolution). Within the \ac{OBPF} passband, we measure the noise floor power, $P_{n}$,  on the OSA at 0.1~nm resolution. We measure the total in-band signal power at 2~nm resolution, denoted as $P_{total}$. In-band signal power is estimated by $P_{sig} = P_{total} - \frac{2}{0.01}\times P_n$. The OSNR is calculated as $10\log_{10}(P_{sig}/P_{n})$. Note that OSNR could vary with frequency due to the channel response. Estimates from this method indicate average OSNR performance over signal bandwidth.

\bmhead{Modulator energy consumption}
Electrical power inside the proposed device is dissipated by charging junction capacitors $C_j$ of the four \acp{MRM} on rising transitions. Assuming that the voltage levels in a given QAM signal are equidistant, the total consumed energy is given by 

\begin{equation}
E_T = 4C_jV^2_{pp} \sum_{i=1}^{\sqrt{N}-1}(\sqrt{N}-i) \left( \frac{i}{\sqrt{N}-1}\right)^2,
\label{eq:Et}
\end{equation}
where, $N$ is the QAM order and $V_{pp}$ is the peak-to-peak voltage swing. Here, the rising transition for a resonator is the falling for the other resonator in the MRA-MZM structure; hence, the total consumed power is multiplied by the number of the \acp{MRM}. Having N total possible transitions and $\log_2(N)$ bit(s) per symbol, the energy consumed per bit is given by 
\begin{equation}
E_b = \frac{E_T}{N\log_2(N)}.
\label{eq:Eb}
\end{equation}

The junction capacitance $C_j$ is estimated to be 18~fF under the operating condition \ifHilight{(see Fig. \ref{fig:exp_res}h)}. Therefore, the effective power consumed for modulation is calculated to be 10.4~fJ/bit for 32QAM and 18.2~fJ/bit for 16QAM. \ifHilight{Figure~\ref{fig:Arch_intro}d compares our work to the state-of-the-art high-bandwidth-density silicon photonic transmitters. Note that only the energy efficiency of modulators is considered; the power consumption of other optical and electronic components should also be considered in system design. As system implementations can vary widely, we confine ourselves to modulators, which are the focus of this study.} 

\section*{Data availability}
The data that supports the findings of this study are available from the corresponding authors upon reasonable request.

\section*{Code availability}
The code used in this study is available from the corresponding authors upon reasonable request.

\section*{Acknowledgments}

This work is funded by NSERC (CRDPJ538381-18). We thank Keysight for the loan of an 80~GHz M8199B \ac{DAC} and 110~GHz UXR RTO.
We thank Nathalie Bacon and Éloi Blouin for their technical support and CMC Microsystems for access to MPW services.

\section*{Author contributions}
W.S. proposed the initial concept and led the project. A.G. and W.S. conceived the chip design. A.G. performed the simulations, designed the mask layout, conducted the optical and d.c. characterization, and designed the stabilization mechanism. S.L. designed the PCB, helped with the stabilization mechanism design, and did the electrical packaging. A.G. performed electro-optic characterization with assistance from Z.Z. F.S. modeled the power-dependent nonlinear behavior of the resonator. Z.Z. and L.A.R designed the transmission experiment. Z.Z. led the data-transmission experiments with assistance from A.G. 
W.S. and L.A.R. supervised the project.

\section*{Competing interests}
The authors declare no competing interests.

\section*{Additional information}
Additional modeling and experimental results are provided in the Supplementary Notes.









\begin{appendices}






\bigskip
\newpage
\bigskip

\begin{Extended Data Table}[h]%
\centering
    \caption{\ifHilightt{Configurations and performance of silicon modulators based on phase modulation in MRMs}}
    \begin{tabular}{lcclllll}
        \toprule[0.1em]
         \multirow{2}*{\textbf{Ref}}&  \multicolumn{2}{c}{\textbf{Configuration}}& \multirow{2}*{\textbf{\shortstack{Detection\\scheme}}}&  \multirow{2}*{\textbf{\shortstack{Modulation\\scheme}}}&  \multirow{2}*{\textbf{\shortstack{Highest \\symbol rate}}}& \multirow{2}*{\textbf{\shortstack{Transmission\\distance}}}\\
         {}& {\textbf{if chirp-free}}& {\textbf{if IQ}}& {}& {}&{} & {}\\
         \midrule[0.1em]
         \cite{Dong2012}& \ding{55} & \ding{51}&  Coherent&  QPSK&  10~GB&  B2B\\
         \midrule[0.05em]
         \cite{Dong2013}& \ding{55}& \ding{51}& Coherent&  QPSK&  28~GB&  B2B\\
         \midrule[0.05em]
         \cite{Romero-Garcia2017}&  \ding{51}& \ding{55} &Direct&  OOK&  30~GB&  B2B\\
         \midrule[0.05em]
         \cite{Li2017hi}&  \ding{51}& \ding{55} &Direct&  PAM4&  46~GB&  Up to 20~km\\
         \midrule[0.05em]
         \cite{Chang2017}&  \ding{51}& \ding{55} & Coherent&  BPSK&  10~GB&  B2B\\
         \midrule[0.05em]
         \cite{DeValicourt2018}&  \ding{55} &\ding{51} & Coherent & QPSK&  20~GB&  Up to 100~km\\
         \midrule[0.05em]         
         \cite{DeValicourt2018}&  \ding{51} &\ding{55} & Direct & OOK&  16~GB&  Up to 20~km\\
         \midrule[0.05em]
         \cite{Jo2024}&  \ding{55} &\ding{51} & Coherent & QPSK&  28~GB&  B2B\\
         \midrule[0.05em]
         \cite{Wu2024}&  \ding{51} &\ding{55} & Direct & PAM4&  112~GB&  Up to 1~km\\
         \midrule[0.05em]
         This work& \ding{51} & \ding{51} &Coherent&  Up to 32QAM&  180~GB&  Up to 80~km\\
         \bottomrule[0.1em]
    \end{tabular}
    \label{tab:comparison}
\end{Extended Data Table}

\bigskip
\newpage
\bigskip

\begin{Extended Data Fig.}[h]%
    \centering
    \includegraphics[width=0.97\textwidth]{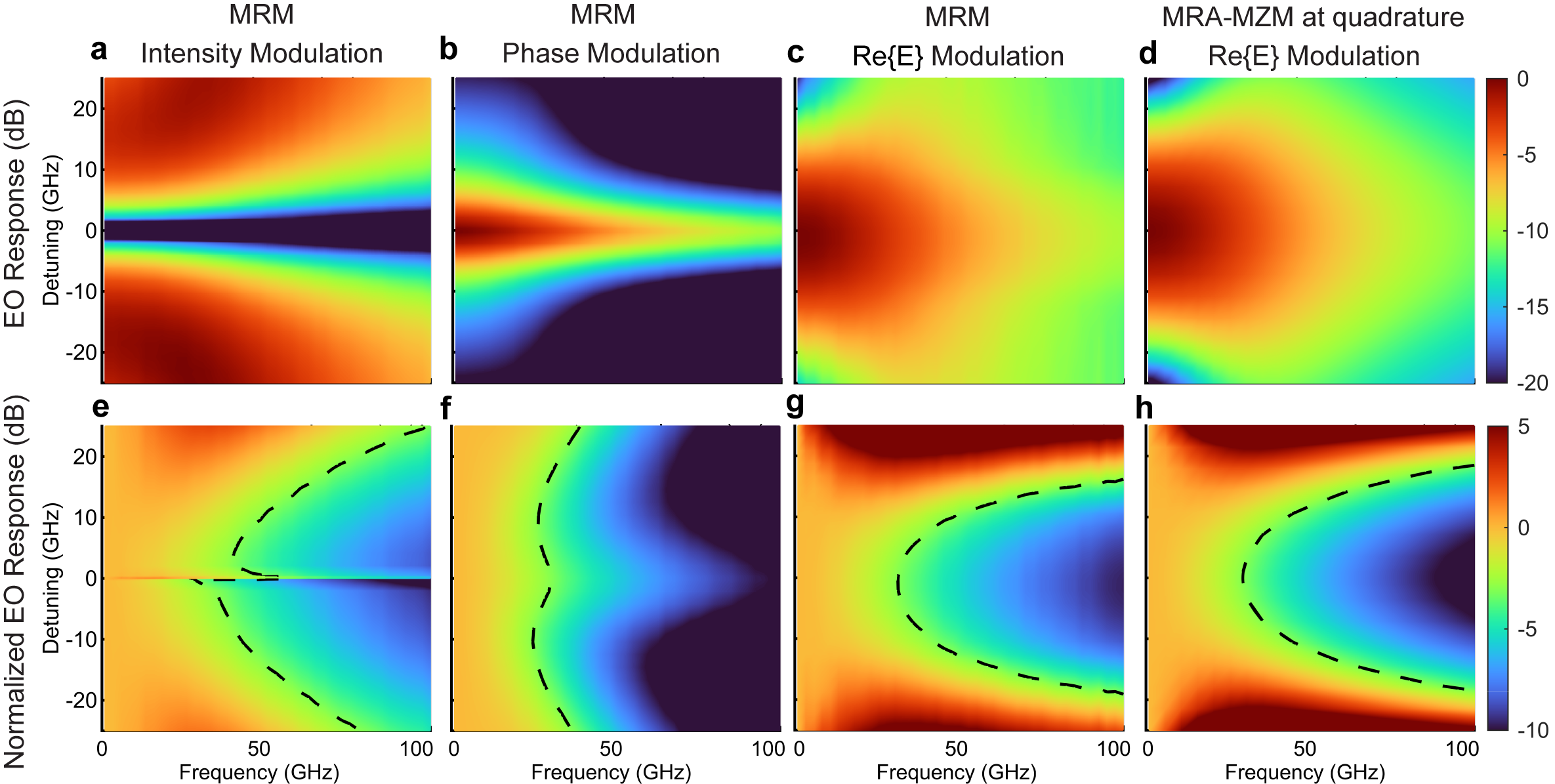}
    \caption{\ifHilightt{\textbf{Small signal electro-optic (EO) response of the modulators under study.} \textbf{a,b,c,d,} Small signal EO intensity modulation (IM) response \textbf{(a)} and phase modulation (PM) responses of MRM\textbf{(b)}, real part of complex amplitude modulation (AM) response of MRM\textbf{(c)}, and real part of complex amplitude modulation \ac{AM} of MRA-MZM \textbf{(d)} normalized to the peak of the graph as a function of the modulation frequency and laser-resonance detuning. \textbf{e,f,g,h,} The EO \ac{IM} \textbf{(e)}, \ac{PM} \textbf{(f)}, real part of \ac{AM} \textbf{(g)}, responses of MRM and real part of \ac{AM} response of MRA-MZM\textbf{(h)} normalized to the low-frequency part for each detuning. Dashed lines indicate 3~dB EO cut-off frequency. 
    Discussions on calculation method are presented in Supplementary Note~\ref{SN:subsec_linear_SS}.}
    }\label{fig:mrm_SSEO_num}
\end{Extended Data Fig.}

\bigskip
\newpage
\bigskip

\begin{Extended Data Fig.}[h]%
\centering
\includegraphics[width=0.9\textwidth]{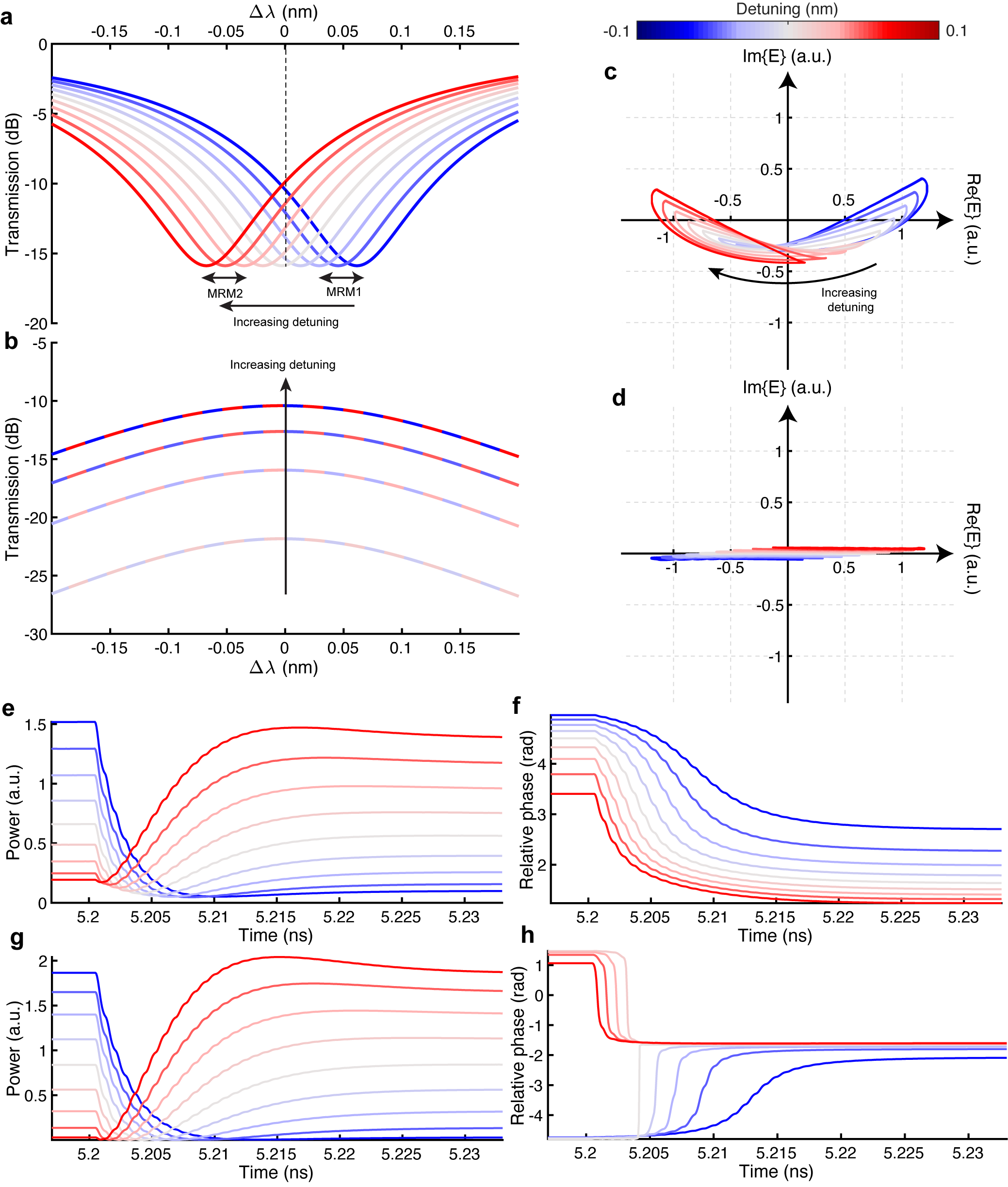}
\caption{\ifHilightt{\textbf{Detuned operation of the modulators.} \textbf{a,b} Transmission spectra of detuned MRMs\textbf{(a)} and MRA-MZMs\textbf{(b)} with zero wavelength offset indicated with dashed lines. The color bar shows the modulation wavelength detuning ($\Delta \lambda_m=\lambda_l - \lambda_r$). \textbf{c,d,} Simulated received data modulated by MRM\textbf{(c)} and MRA-MZM\textbf{(d)} with a \ac{PAM}-2 signal across a spectrum of detunings, collectively presented on the complex plane.  
The MRM trajectories do not intersect the origin, in contrast to the MRA-MZM, indicating a limited \ac{ER}. 
Power and relative phase of MRM \textbf{(e,f)} and MRA-MZM \textbf{(g,h)} of the received signal over time for different laser-resonance detunings with an evident enhancement of \ac{ER} in MRA-MZM.
} 
\ifHilightt{Observations reveal that the received signal from MRM adheres to the anticipated trend, distancing itself from the origin. Intriguingly, the transition between the two states becomes more intricate, resulting in a circular trajectory. In contrast, the MRA-MZM trajectory remains straight. This behavior aligns with expectations derived from the spectral response.}  
}\label{fig:mrm_operate}
\end{Extended Data Fig.}

\bigskip
\newpage
\bigskip

\begin{Extended Data Fig.}[h]%
\centering
\includegraphics[width=0.99\textwidth]{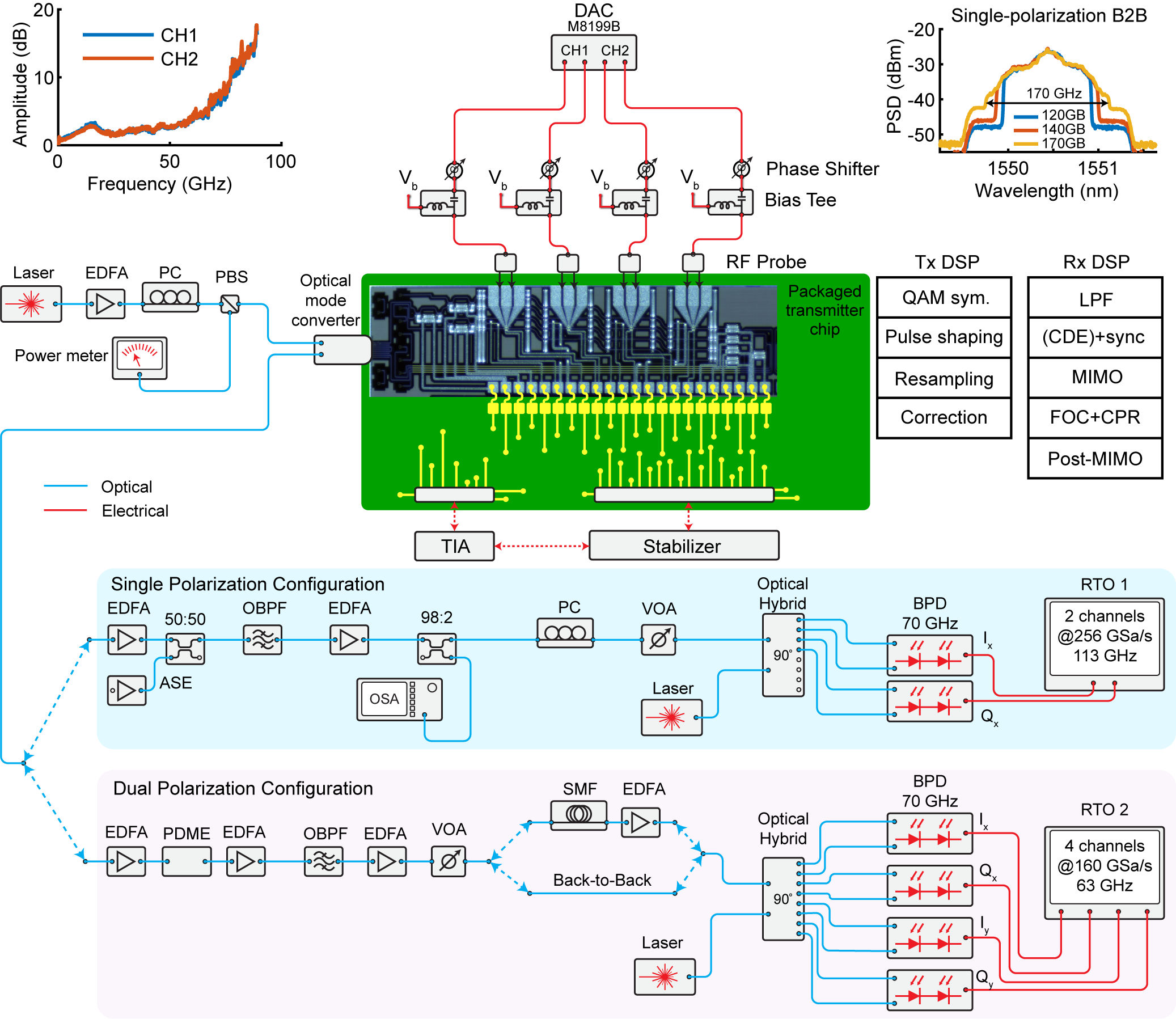}
\caption{\textbf{Data transmission experiment setup.} Experimental setup of single polarization (SP) and dual polarization (DP) transmission with coherent detection. At top, from left to right and top to bottom, are: (1) waveform correction; (2) optical spectrum at various symbol rates; and (3) DSP stacks for transmitter and receiver. DAC: digital-to-analog converter, EDFA:erbium-doped fibre amplifier, PC: polarization controller, PBS: polarization beam splitter, TIA: trans-impedance amplifier, ASE: amplified spontaneous emission, OBPF: optical bandpass filter, VOA: variable optical attenuator, BPD: balanced photodetector, RTO: real time oscilloscope, PDME: polarization division multiplexing emulator, SMF: single mode fiber, DSP: digital signal processing, PSD: power spectral density, B2B: back-to-back, Tx: transmitter, Rx: receiver.
}\label{fig:trans_exp_setup}
\end{Extended Data Fig.}

\bigskip
\newpage
\bigskip

\begin{Extended Data Fig.}[h]%
\centering
\includegraphics[width=0.69\textwidth]{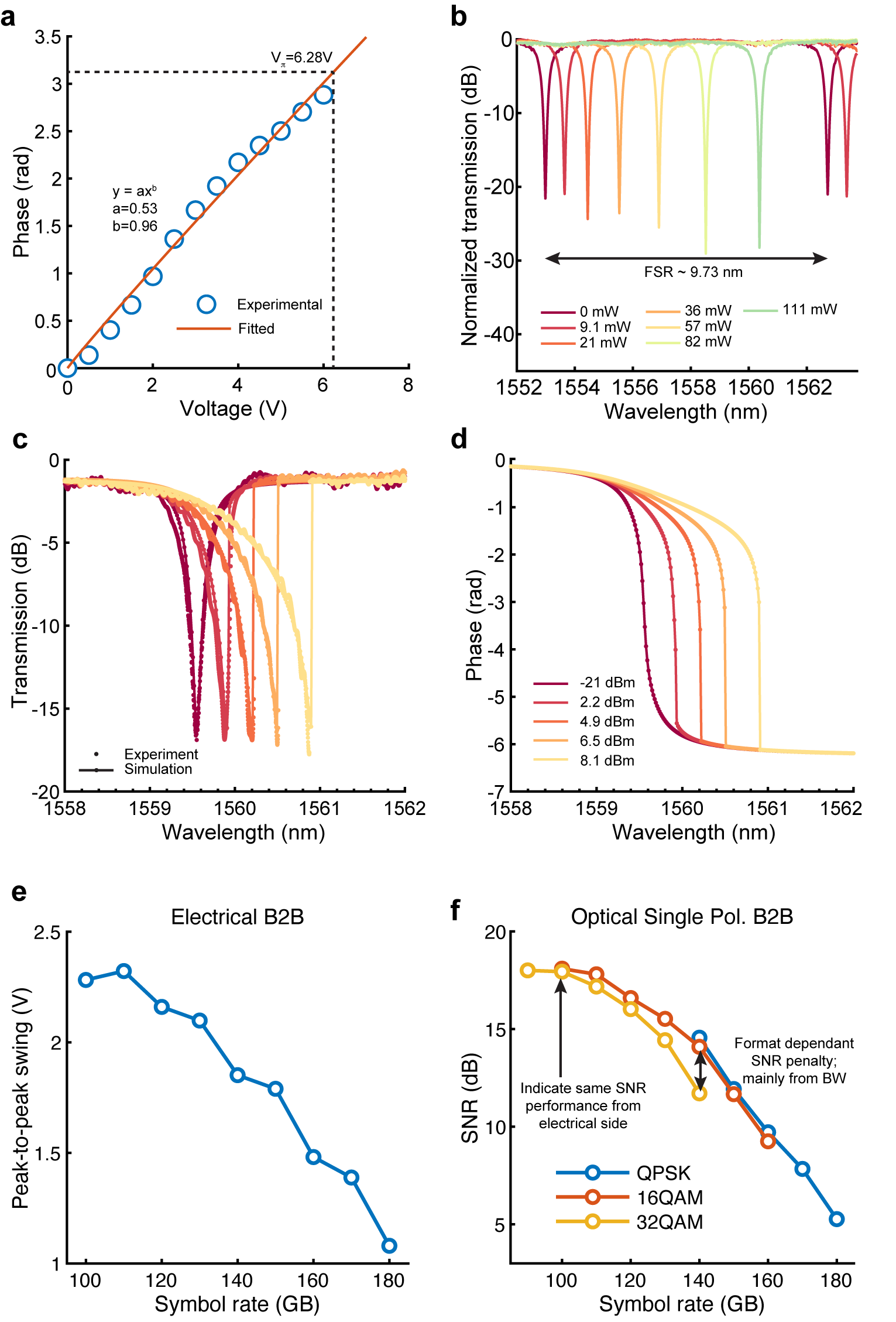}
\caption{\ifHilight{\textbf{Extended experimental characterization results.} \textbf{a,} Phase shift of the MRM output as a function of applied voltage, showing a $V_\pi$ of 6.28~V. \textbf{b,} MRM heater characterization result showing a half-wave power (P\textsubscript{\textpi}) of $\sim 60~mW$. Note that the transmission is normalized to the maximum. \textbf{c,d,} Power-dependent measurement (solid) and simulated (dashed) transmission and phase spectra of \ac{MRM}. \textbf{e, }Electrical signal swing versus symbol rate, estimated from a PAM4 signal with 2.7~V swing setting on Keysight M8199B. The swing decreases with stronger compensation. \textbf{f, }SNR estimated from recovered constellations after optical transmission (single polarization, B2B case). At a symbol rate of 100 Gbaud, we observe the same SNR across QPSK to 32QAM formats, indicating a very similar driving swing from the AWG with varied formats. We attribute the SNR penalty of 32QAM at higher rates to its higher sensitivity to channel bandwidth limitations.
}
}\label{fig:exp_res_2}
\end{Extended Data Fig.}

\bigskip
\newpage
\bigskip

\newpage
\section*{Supplementary Note}

\pagenumbering{arabic} 
\setcounter{page}{1} 

\startcontents[Supplementary Note]
\printcontents[Supplementary Note]{}{1}{}
\bigskip
\newpage
\bigskip

\section{Supplementary Note: Coupling regimes in MRM}\label{SN:sec_MRM_coupling}

In an all-pass microresonator comprised of a ring coupled to a bus waveguide, which is presented in Fig.~\ref{fig:SN_MRM_coupling}a, assuming the single mode condition for waveguide and steady-state continues wave operation, the electric field transfer function can be written as \citeS{Heebner2008}:

\begin{equation}
\label{eq:SN_MRM_ETF}
\frac{E_o}{E_i} = e^{i(\pi+\phi)} \frac{a - \sigma e^{-i\phi}}{1 - \sigma ae^{i\phi}}
\end{equation}
\noindent where $\phi$, $a$, and $\sigma$ are round-trip phase shift, amplitude transmission, and electric field through coupling coefficient, respectively. In a lossless coupler, the cross-coupling ($\kappa$) and through-coupling ($\sigma$) coefficients are related through $\kappa^2 + \sigma^2 = 1$. The intrinsic decay inside the resonator ($1 - a^2$) arises from various factors, such as bending loss, absorption, and scattering within the resonator.

\ifHilight{The coupling between the MRM and the bus waveguide can be classified into three conditions: under-coupled, critically-coupled, and over-coupled. Under the critical coupling condition, the coupled power to the resonator equals the power lost over a roundtrip inside the resonator ($\kappa^2 = 1 - a^2$), and the output power of the resonator reaches zero. Under-coupling and over-coupling conditions refer to the scenarios in which the bus-resonator coupling is weaker ($\kappa^2 < 1 - a^2$) and stronger ($\kappa^2 > 1 - a^2$) than the intrinsic decay rate of the microresonator, respectively.}

\ifHilight{Phasor diagrams of the normalized electric field transfer function are intuitive representations of both the amplitude and phase of a resonator. The transmission spectra of three identical MRMs with different coupling conditions (i.e., under-coupled, lightly over-coupled, and highly over-coupled) are depicted in Figs.~\ref{fig:SN_MRM_coupling}b-d using phasor diagrams and conventional transmission and phase diagrams. One can notice that the phase response of the MRM falls into three categories, as presented in Fig.~\ref{fig:SN_MRM_coupling}b and d. The phase response of an MRM strongly depends on the coupling condition.}

\ifHilight{Figure~\ref{fig:SN_MRM_coupling}e shows the electric field transfer function for $\kappa^2$ within the $2.9\% - 29\%$ range in 10 steps with equal spacing on the complex plane. In this example, with fixed intracavity loss, the power coupling coefficient ($\kappa^2$) equal to $8.7\%$ corresponds to the critical coupling condition. Therefore, a coupling coefficient below or above this value results in under-coupling or over-coupling conditions, respectively. In this case, the line color represents the wavelength detuning in nm.}

\begin{figure}[th]%
\centering
\includegraphics[width=0.99\textwidth]{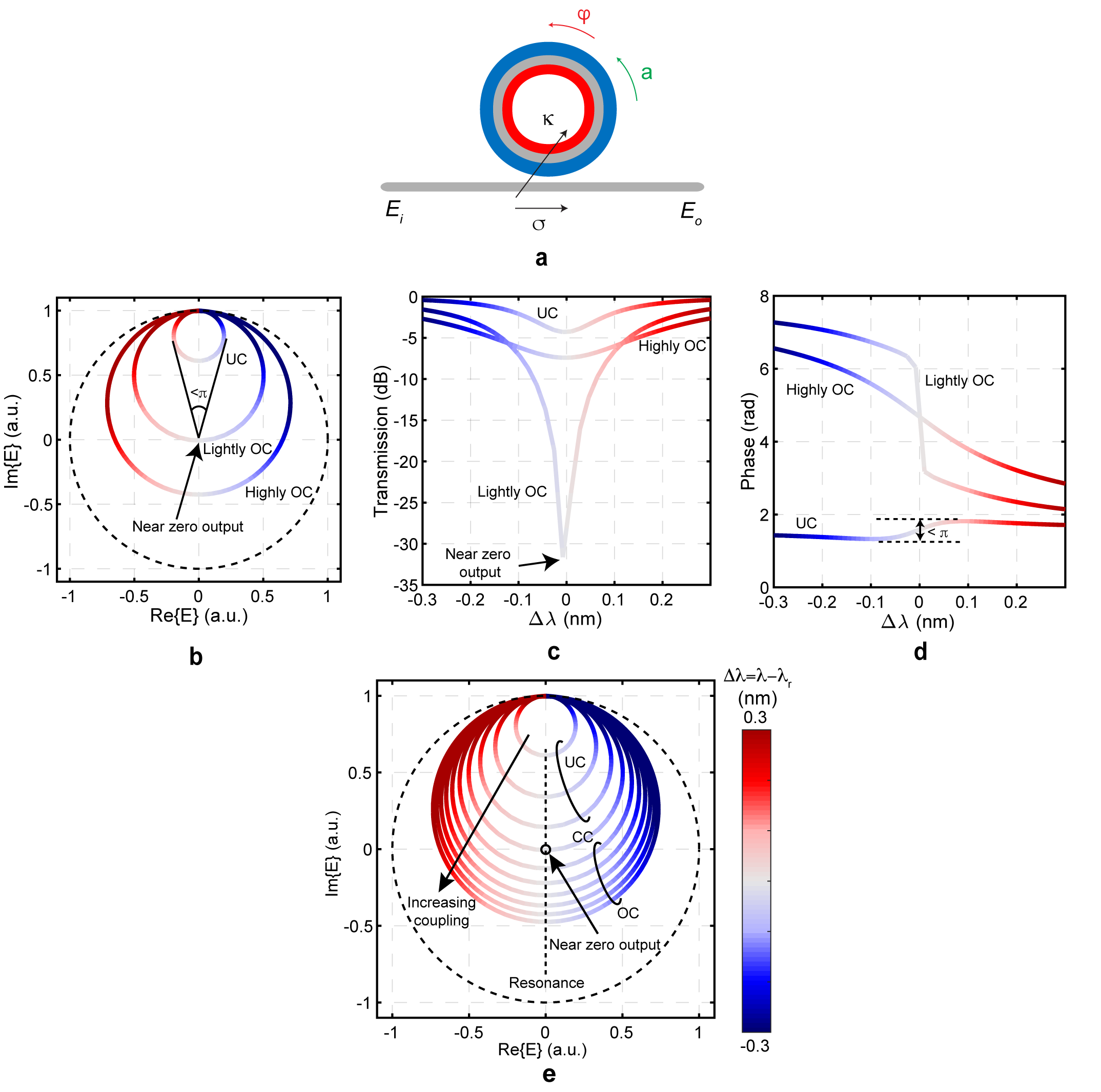}
\caption{\ifHilight{\textbf{(a),} Schematic of a MRM in an all-pass configuration. \textbf{(b),(c),(d)} Transmission spectrum as a function of wavelength of the MRM in under-coupled, lightly over-coupled (near critical coupling), and highly over-coupled conditions in a phasor diagram (b), and conventional transmission (c) and phase (d) spectra. \textbf{(e),} Phasor diagram of the MRM for various coupling coefficients.}
}\label{fig:SN_MRM_coupling}
\end{figure}

One can notice from the inequal spacing between phasor lines in Fig.~\ref{fig:SN_MRM_coupling}e that the sensitivity of the transfer function is higher for undercoupling compared to overcoupled MRM. This means that a small perturbation in $\kappa$ that can potentially be caused by the fabrication process variation may lead to a large variation in the response of the fabricated device and a big contrast between the designed and fabricated device. 


\pagebreak
\bigskip
\newpage
\bigskip

\section{Supplementary Note: Microring modulator modeling}\label{SN:sec_MRM_model2}

\subsection{Small-signal modeling of MRM}\label{SN:subsec_MRM_model_ss}
In this section, we study the behavior of a single-ring MRM using a small-signal model. Under the small-signal assumptions, the total transfer function of a ring modulator, $H(\omega_m, \Delta\omega)$, is defined as: 
\begin{equation}
H(\omega_m, \Delta\omega) = \frac{\hat{P}_{out}}{\hat{V}_{in}},
\label{eq:trans}
\end{equation}
\noindent 
where $\hat{P}_{out}$ is a complex small-signal phasor of the output power in the through port of a single ring and $\hat{V}_{in}$ is the small-signal phasor which is equal to the amplitude of input voltage with modulation angular frequency of $\omega_m$. Angular frequency detuning of resonance frequency and input laser frequency is defined as $\Delta\omega = \omega_{in} - \omega_{r} $. 

The transfer function in Eq.~\ref{eq:trans} is calculated from three transfer functions, namely electrical ($H_E$), electro-optical ($H_{EO}$), and optical ($H_O$) using $H(\omega_m) = H_E(\omega_m) \times H_{EO} \times H_O(\omega_m, \Delta\omega)$ (Fig. \ref{fig:SN_mrm_model_ss}a) written in s-domain as \citeS{Karimelahi2016}:

\begin{equation}
H(s) = G_{DC}\frac{1}{1+\frac{s}{\frac{1}{R_{eq}C_{eq}}}} \left[ \frac{(-\frac{s}{z}+1)\omega_n^2}{s^2 + 2\zeta \omega_n s + \omega_n^2} \right].
\label{eq:trans_s}
\end{equation}

\noindent where $\omega_n = \sqrt{\Delta\omega^2 + 1}$, $\zeta = 1 / \sqrt{1 + (\tau\Delta\omega)^2}$, $R_{eq}$, and $C_{eq}$ are the equivalent resistance and capacitance of the dominant pole of the electrical transfer function ($H_E$) which is a modeled as a low pass filter.
The DC gain ,$G_{DC}$, is:

\begin{equation}
G_{DC} = \frac{2\mu^2 P_{in} (\frac{\omega_r}{n_g} \frac{\partial n_{eff}}{\partial v} \mid _{V_{DC}})}{\omega_n^4} 
\left( 
2 \frac{\Delta\omega}{\tau_i} + \tan(\phi_{H_{EO}}) 
\left( \Delta\omega^2 - \frac{1}{\tau}(\frac{1}{\tau_i}-\frac{1}{\tau_e})\right)
\right)
\label{eq:GDC}
\end{equation}

\noindent where $n_g$ is the group index, $n_{eff}$ is the effective refractive index, $P_{in}$ is the input power, and $\tau$ is the electric field amplitude decay time constant. $\tau$ is calculated using $1/\tau = 1/\tau_e +1/\tau_i$, with $\tau_e$ and $\tau_i$ being the amplitude decay time constants due to the ring-bus waveguide cross-coupling and intrinsic loss inside the cavity, respectively. Here, 
$\mu$ is the mutual ring-bus cross-coupling coefficient, which is related to $\tau_e$ through $\mu^2 = 2/\tau_e$ and $\phi_{H_{EO}}$ is the phase of electro-optical transfer function ($H_{EO}$).

The transfer function in Eq.~\ref{eq:trans_s} shows that MRM is a third-order system with one real zero, one real pole, and a complex conjugate pole pair at the following locations:
\begin{gather}
z = \frac{ 2 \frac{\Delta\omega}{\tau_i} + \tan(\phi_{H_{EO}}) \left( \Delta\omega^2 - \frac{1}{\tau}(\frac{1}{\tau_i}-\frac{1}{\tau_e})\right), }{ \tan(\phi_{H_{EO}})(\frac{1}{\tau_i}-\frac{1}{\tau_e}) - \Delta\omega }\nonumber \\
p_{1,2} = -\frac{1}{\tau}\pm j\Delta\omega, \nonumber \\
p_3 = -\frac{1}{R_{eq}C_{eq}}.
\label{eq:polzer}
\end{gather}

Locations of the zero and poles on the complex plane are presented in Fig.~\ref{fig:SN_mrm_model_ss}b. As it is depicted in Fig.~\ref{fig:SN_mrm_model_ss}b and Eq.~\ref{eq:polzer}, $G_{DC}$ and the location of the $z$ is a function of coupling condition, photon lifetime, and frequency detuning. Fig.~\ref{fig:SN_mrm_model_ss}c shows the location of the zero in the vicinity of zero detuning normalized to $-2/\tau_i$ for an undercoupled ($\tau_e > \tau_i$) and an overcoupled ($\tau_i > \tau_e$) MRM. As it is shown, the location of the zero experiences a discontinuity as it gets close to the zero detuning. The offset between zero detuning and the detuning that the discontinuity occurs depends on the coupling condition and is the opposite for uncercoupled and overcoupled MRMs. Moreover, the location of the zero depends on the sign of the detuning, which has an impact on the total transfer function of the MRM, leading to asymmetric side-band generation in detuned MRMs \citeS{Muller2015}. Sign of $G_{DC}$ changes across zero detuning, and the peak on the negative detuning side is stronger, as illustrated in Fig.~\ref{fig:SN_mrm_model_ss}d. According to Eq.~\ref{eq:polzer}, the real and imaginary parts of $p_{1,2}$ depend on the photon lifetime inside the resonator and the detuning, respectively. 

At zero detuning ($\Delta\omega = 0$), according to Eq.~\ref{eq:polzer} imaginary part of the $p_{1,2}$ equals to zero, and the real part of it coincides with the location of the zero resulting in cancellation of a pole with the zero from the overall system, reducing the system to a second order system with two real poles. Non-zero small signal modulation of the MRM at resonance wavelength can be attributed to the non-zero $G_{DC}$ at zero detuning resulting from simultaneous index and loss modulation.

\begin{figure}[]%
\centering
\includegraphics[width=0.99\textwidth]{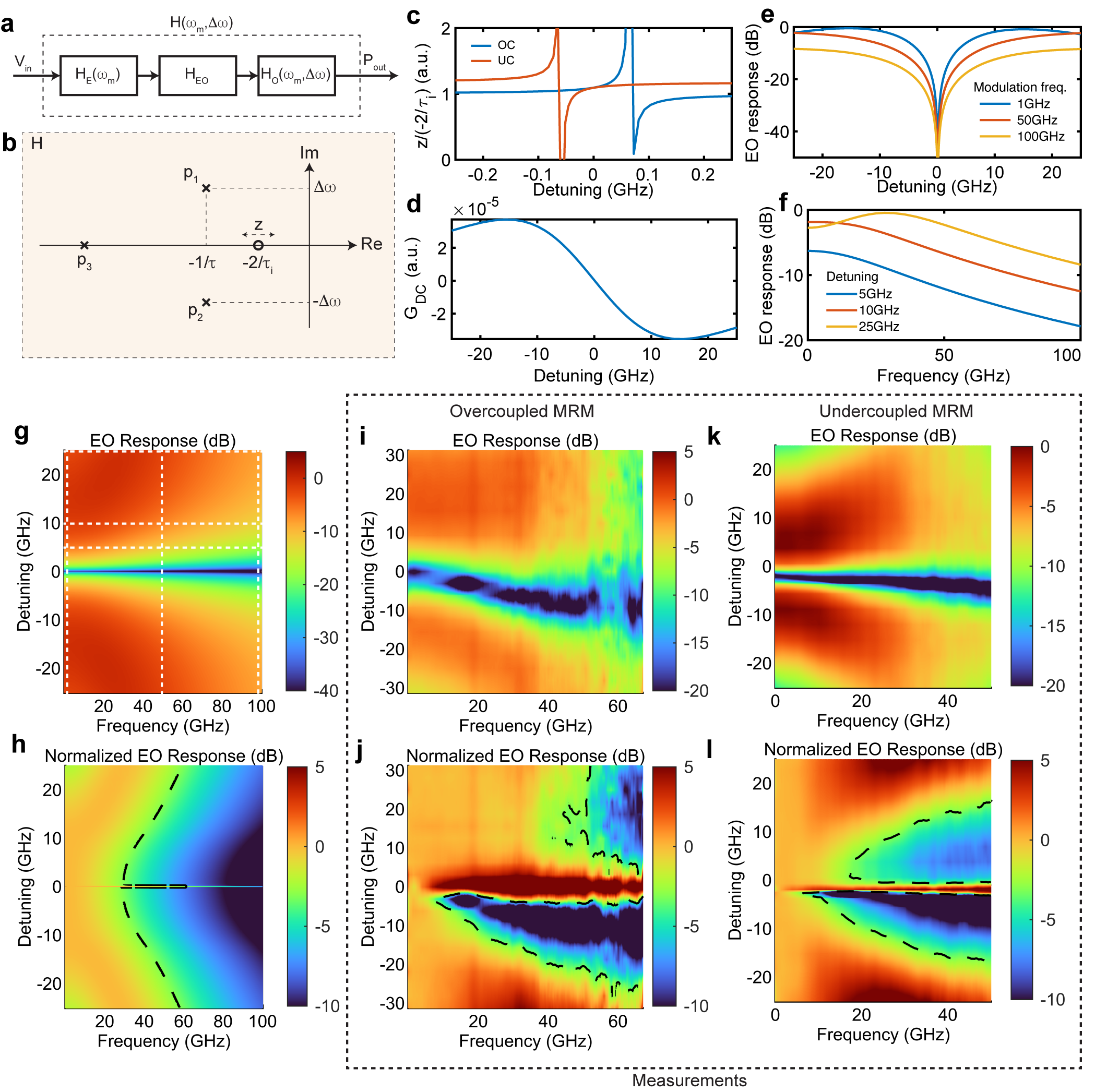}
\caption{\textbf{Small-signal modeling of MRM} \textbf{a,} Transfer function of MRM comprised of three cascaded transfer functions. \textbf{b,} Pole-zero representation of the MRM transfer function. \textbf{c,} location of the zero in overcoupled and undercoupled MRM. \textbf{d,} The DC gain as a function of detuning frequency. \textbf{e,f,} MRM EO response vs detunig \textbf{(e)} and modulation frequency \textbf{(f)} for dashed lines indicated in \textbf{(g)}. \textbf{g,h,} OMA \textbf{(g)} and normalized EO response \textbf{(h)} of MRM vs detuning and modulation frequency. \textbf{i,j,k,l,} measured small-signal OMA and EO response of an overcoupled \textbf{(i-j)} and an undercoupled \textbf{(k-l)} MRM. 
}\label{fig:SN_mrm_model_ss}
\end{figure}

Using the small signal transfer function, we can plot the electro-optic response for different modulation frequencies across a range of detunings (see Fig.~\ref{fig:SN_mrm_model_ss}e). The amplitude of the \ac{IM} peaks on two sides of the resonance wavelength and deeps on zero detuning, which was predicted considering the gradient of the static transfer function of the MRM. Also, the electro-optic response for various detunings is plotted in Fig.~\ref{fig:SN_mrm_model_ss}f for a range of modulation frequencies where optical peaking is observed. 

To further study the interplay of the contributing factors to the EO response and the EO bandwidth of MRM, we have plotted the EO response normalized to its maximum representing the OMA in Fig.~\ref{fig:SN_mrm_model_ss}g and normalized EO response normalized to the lowest frequency component of the EO response in Fig.~\ref{fig:SN_mrm_model_ss}h. The white dashed lines on Fig.~\ref{fig:SN_mrm_model_ss}g indicates the cross-sections plotted in Fig.~\ref{fig:SN_mrm_model_ss}e and f. The small-signal normalized EO response decreases as the modulation frequency increases, which is clearly observed in the normalized EO response plot Fig.~\ref{fig:SN_mrm_model_ss}h. The dashed line shows the 3-dB cut-off frequency, which enhances at a higher detuning frequency due to the peaking effect. We performed small-signal EO response measurements on an overcoupled and an undercoupled MRM to qualitatively confirm the behavior observed from the small-signal model. The measurement results are presented in Fig.~\ref{fig:SN_mrm_model_ss}i-l. 
Measured normalized EO responses are also showing the discontinuity around the zero detuning as predicted by model.

\subsection{Numerical simulations of small-signal electro-optic response of MRM}\label{SN:subsec_linear_SS}

\ifHilightt{In our small-signal analysis, the complex electric field is captured, from which the real part of complex amplitude, intensity, and phase modulation transfer functions, $H_{\text{MRM, AM}}=\frac{\Delta Re\{ E_{out}\}}{\Delta V_{in}}$ , $H_{\text{MRM, IM}}=\frac{\Delta P_{out}}{\Delta V_{in}}$ and $H_{\text{MRM, PM}}=\frac{\Delta \Phi_{out}}{\Delta V_{in}}$, are extracted, as illustrated in Fig.~{\ref{fig:SN_mrm_model_linear_ss}}~a, where $f_m$ represents the modulation frequency. To observe the linear portion of the responses, a band-pass filter centered at the modulation frequency can be applied, which removes higher-order harmonics, analogous to the process of measuring S-parameters with a VNA in a linear configuration.}

\ifHilightt{IM near the resonance wavelength (zero detuning) creates higher-order harmonics with square-law detection. 
This leads to different frequency responses with and without using the band-pass filter. Figure~\ref{fig:SN_mrm_model_linear_ss}~b shows the normalized IM response of the MRM, simulated with and without the aforementioned band-pass filter. A strong peaking appears in the linear response (with the filter), which agrees well with experimental observations illustrated in Fig.~\ref{fig:exp_res}~i. The strong peaking effect at zero detuning in the MRM response arises from a complex coupling between amplitude and phase, manifesting as a filtering effect when detected using a linear VNA was also observed in others' work \cite{Yu2014}. Our numerical simulation agrees well with the experimental results in Fig.~\ref{fig:exp_res}~i.}

\ifHilightt{A similar behavior occurs in IM response of the MRA-MZM operating at the null point of the \ac{MZI} when detected with a square-law detector. However, none or little higher-order harmonic generation is expected when the device is operated at the quadrature point of the \ac{MZI}, where IM shows the best linearity. 
Note that as the phase modulation is stronger than intensity modulation at resonance wavelength, the frequency response doesn't show a strong peaking, even though they remain biased at zero detuning.
To confirm this, we have plotted the normalized IM response of the MRA-MZM at the quadrature point, with and without using the band-pass filter, in Fig.~\ref{fig:SN_mrm_model_linear_ss}~c. The close resemblance of the two curves confirms that the response captured with the square-law detector does not introduce significant nonlinear behavior. To provide a comprehensive view of the coherent dynamics of the devices without the artifacts created by the detection mechanism, the frequency responses of IM, PM, and real component of complex amplitude for MRM and MRA-MZM calculated without using the aforementioned band-pass filter for a range of laser-resonance detunings, as presented in Extended Data Fig.~\ref{fig:mrm_SSEO_num}. Note that the imaginary component of complex amplitude for MRA-MZM is negligible as the trajectory on the complex plane is straight.}

\begin{figure}[t]%
\centering
\includegraphics[width=0.99\textwidth]{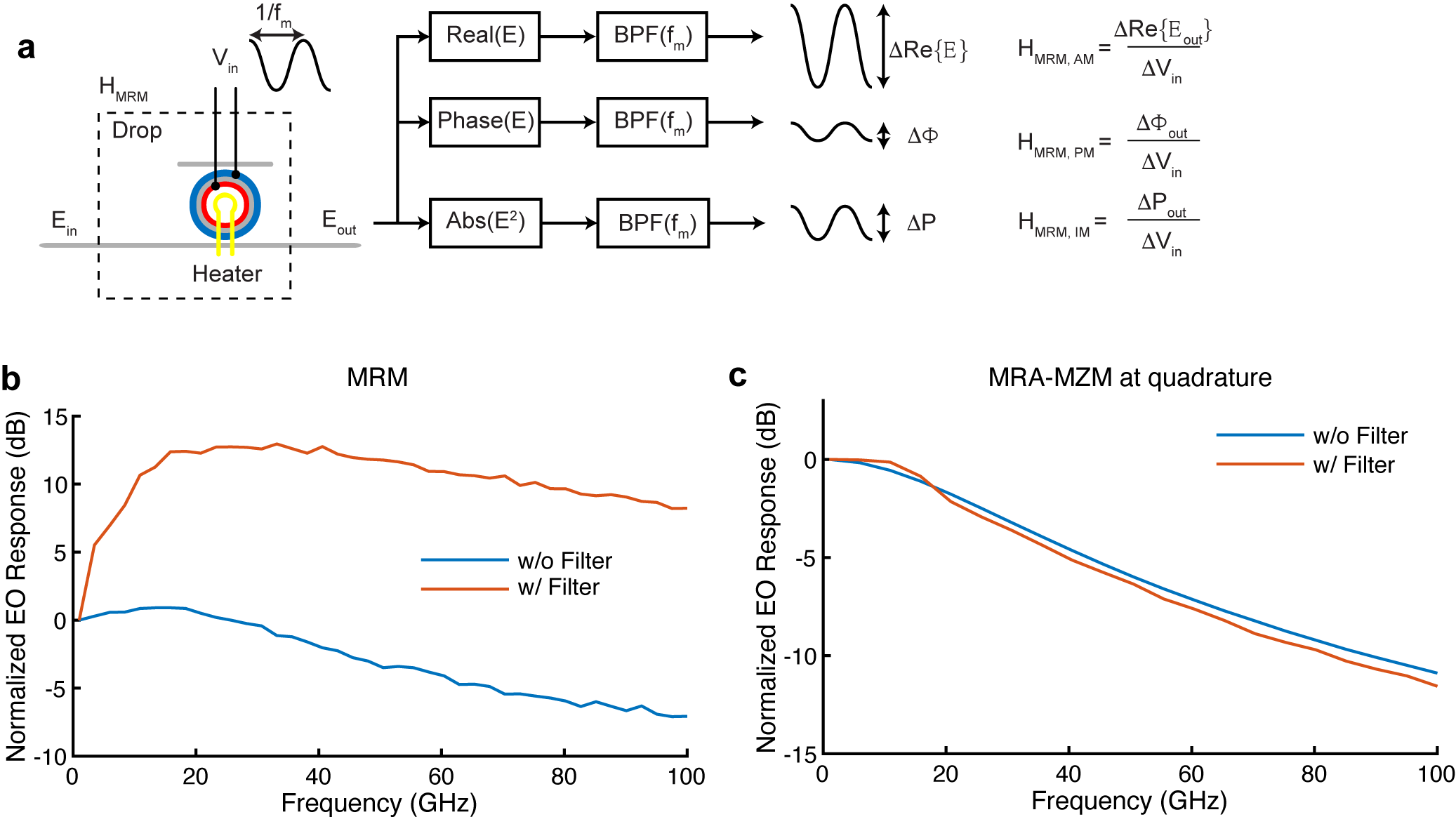}
\caption{\ifHilightt{\textbf{Numerical simulations of the small-signal linear electro-optic response of MRM} \textbf{a, }extraction of small-signal amplitude, intensity and phase modulation transfer function of MRM. The presence of a bandpass filter centered at the modulation frequency results in capturing only the linear part of the EO response. \textbf{b, } Comparison of the IM response of MRM at resonance wavelength with and without bandpass filter. \textbf{c, } Comparison of the IM response of MRA-MZM at quadrature point while detuning is set to zero with and without bandpass filter.   } 
}\label{fig:SN_mrm_model_linear_ss}
\end{figure}

\subsection{Optical nonlinear modeling of MRM response }\label{SN:subsec_power_sen}

When the optical power increases inside the ring, optical nonlinear effects start to change the behaviour of the system. These nonlinear effects as well as self-heating could be analyzed by a set of coupled differential equations governing the time evolution of temperature, concentration of carriers, and the optical field. A comprehensive analysis of these coupled differential equations has been carried out in \citeS{Johnson}.

In a general term, the transmission of an MRM could be written as \citeS{Poon}:
\begin{equation}\label{eq:SN_pow_1}
T(t)=\sigma+a(t) e^{-j \phi(t)}[\sigma T(t-\tau)-1],
\end{equation}
where $\sigma$, $a(t)$, $\phi(t)$, and $T(t)$ are, respectively, the field through coupling coefficient, time-dependent round-trip amplitude transmission, round-trip phase shift, and complex transmission. Note that Eq. \ref{eq:SN_pow_1} accounts for all-pass structure; however, we can consider the loss from the weakly coupled drop port as a distributed loss and add it to the other loss sources in an all-pass configuration and proceed with Eq. \ref{eq:SN_pow_1}.  Since $a(t)$ and $\phi(t)$ are general, we can encapsulate all the time-dependent loss and phase shifts due to either modulation or nonlinear effects into them. Therefore, the next step is to find $a(t)$ and $\phi(t)$ in presence of nonlinear effects. 

Two-photon absorption (TPA) is the primary source of nonlinear loss. As a result of this process, two photons are absorbed and create an electron-hole pair. Therefore, the first nonlinear loss is related to TPA. The generated electron-hole pairs will introduce another source of loss by free carrier absorption (FCA) \citeS{Ram}. The overall 
$a(t)$ could be written as $a(t) = a_0 a_v(t) a_{TPA}(t) a_{FCA}(t)$, where $a_0$ and $a_v(t)$ are round-trip amplitude transmission due to background loss and modulation loss, and $a_{TPA}(t)$, $a_{FCA}(t)$ are amplitude transmissions due to the loss terms associated with TPA and FCA, respectively. On the other hand, the 
phase shift is $\phi(t) = \phi_0 +\phi_v(t)+ \phi_{FCD}(t) + \phi_{Th}(t)$ where $\phi_{Th}$ is the phase shift induced by generated heat. 
$\phi_0$ is the passive round-trip phase shift in the ring, and $\phi_v(t)$ is the modulation phase shift due to the applied voltage to the junction. $\phi_{FCD}$ is the phase shift introduced by the free carriers that were generated via TPA. In order to define these nonlinear phase shifts and loss terms, we need to know the time-dependent carrier concentration, temperature, and energy density inside the ring. The dynamics of carriers and temperature are governed by the two following equations \citeS{LipsonSM, Shateri2024}.

\begin{equation}\label{eq:SN_pow_2}
\frac{\mathrm{d} N(t)}{\mathrm{~d} t}=-\frac{N(t)}{\tau_c}+\frac{P_A(t)}{2 h {\nu} V } \quad \text{and} \quad \frac{\mathrm{d}  {\theta}(t)}{\mathrm{~d} t}=-\frac{ {\theta}(t)}{\tau_{th}}-\frac{P_A(t)}{\rho C V},
\end{equation}

\noindent where $N$ and $\theta$ are, respectively, the density of generated free carriers and temperature, $\tau_c$ is free carrier time constant, V is the ring's effective volume, $\tau_{th}$ is the thermal time constant, and $\rho$ and $C$ are, respectively, the density of silicon and the thermal heat capacity of silicon. $h$ and $\nu$ are Plank's constant and the frequency. $P_A$ is the total absorbed power. 

Since the total absorbed power is determined by $a(t)$, and the absorbed power acts as a source in Eq.~\ref{eq:SN_pow_2} the three parameters $N(t)$, ${\theta}(t)$, and $T(t)$ are mutually coupled. In order to solve the coupled equations, at each round-trip, we first calculate the field amplitude and phase inside the ring and then update the values of the $\theta$ and $N$. On the other hand, the modified temperature and carrier concentration, affect the field in the next round trips and so on. 

After solving for $N(t)$ and $\theta(t)$ for each round-trips, we can calculate $\phi_{FCD}(t)$, $\phi_{Th}(t)$, $a_{TPA}(t)$, and $a_{FCD}(t)$. Since we have the total $a$ and $\phi$ specific to each round-trip, we can use Eq.~\ref{eq:SN_pow_1} to obtain the time domain transmission of the system. Note that if there were no nonlinear effects, the round-trip amplitude transmission and the phase factor remained the same for each round-trip.
\begin{figure}[h]%
\centering
\includegraphics[width=0.99\textwidth]{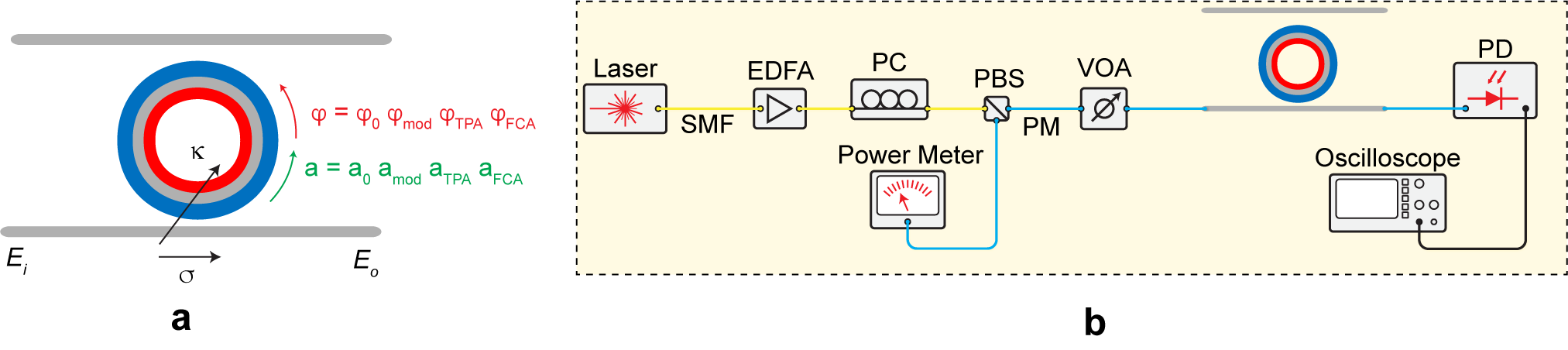}
\caption{\textbf{Nonlinear modeling of MRM response. }\textbf{a,} Schematic of an MRM and the decomposition of the total round-trip phase and amplitude transmission into linear and nonlinear parts. \textbf{b,} Measurement setup used to extract the transmission response of the MRM as a function of the input optical power.}\label{fig:SN_power}
\end{figure}

To verify our model, we experimentally characterized an MRM loaded with PN junction identical to the MRM used in the prototype at various levels of input optical powers. The schematic diagram of the setup is depicted in Fig.~\ref{fig:SN_power}b. The measurement results are provided in \ifHilight{Extended Data Fig.~\ref{fig:exp_res_2}c and d}.


\section{Supplementary Note: S-parameter characterization }\label{SN:sec_EO_char}

S-parameter characteristics of the devices investigated in this work were measured with a 67~GHz \ac{VNA} (Keysight PNA-N5227A). \ifHilightt{
In the S-parameter measurements, the calibration reference plane is established at the contact point between the probe and the device under test (DUT) using proper de-embedding procedures integrated into the VNA (Keysight PNA-N5227A). This ensures that the responses presented in this work are not significantly influenced by components of the measurement setup prior to the probe-DUT contact point (i.e., the calibration reference plane), including the 50~$\Omega$ termination installed on the probe.}

The EO S\textsubscript{21} response of the MRM and MRA-MZM was then captured by the 70-GHz PD (Finisar BPDV3120). The frequency response of the photodiode was obtained from its impulse response, and it was used to compensate for the measured S\textsubscript{21} data to achieve the frequency response of only the MRM at a bias of 2~V. Figure \ref{fig:SN_eo_char} shows the measurement setups to extract the S-parameters. \ifHilightt{Measurement averaging is performed with approximately 500 averages in each case, reducing the noise floor of the measurement setup by $13~dB$. The noise is generally uncorrelated between successive measurements.}

The amplitude of the narrow band laser is attenuated via a voltage-controlled VOA to ensure that the effect of nonlinear phenomena that can lead to asymmetric resonance shape and electro-optic bandwidth extension is minimized. The laser wavelength is set to the blue side of the resonance wavelength and swept toward longer wavelengths, and scattering parameters are recorded in each step. An \ac{EDFA} is used to amplify the modulated optical signal coming from MRM and passed through an optical passband filter to minimize the noise received at the receiver.

To capture the scattering parameters for MRA-MZM, which is a dual drive modulator, we used a balun to generate differential signals from the output of 2-port VNA. Using phase matched cables we minimize the phase difference between two signals. Finally, phase difference between these two signals is tuned using a pair of RF phase shifters. Alignment of the resonance wavelength of the two MRMs is done by using a stabilizer developed for this purpose. The resonance wavelength of the MRMs are automatically aligned to the laser wavelength ensuring the operation at zero detuning. Further details about the stabilization system is presented in Supplementary Note~\ref{SN:sec_stab}. \ifHilightt{Using the integrated thermo-optic phase shifter integrated on the MZM, the phase difference is set to \text{$\pi/2$} (i.e., the quadrature point) for S-parameter characterization. }

\begin{figure}[th]%
\centering
\includegraphics[width=0.89\textwidth]{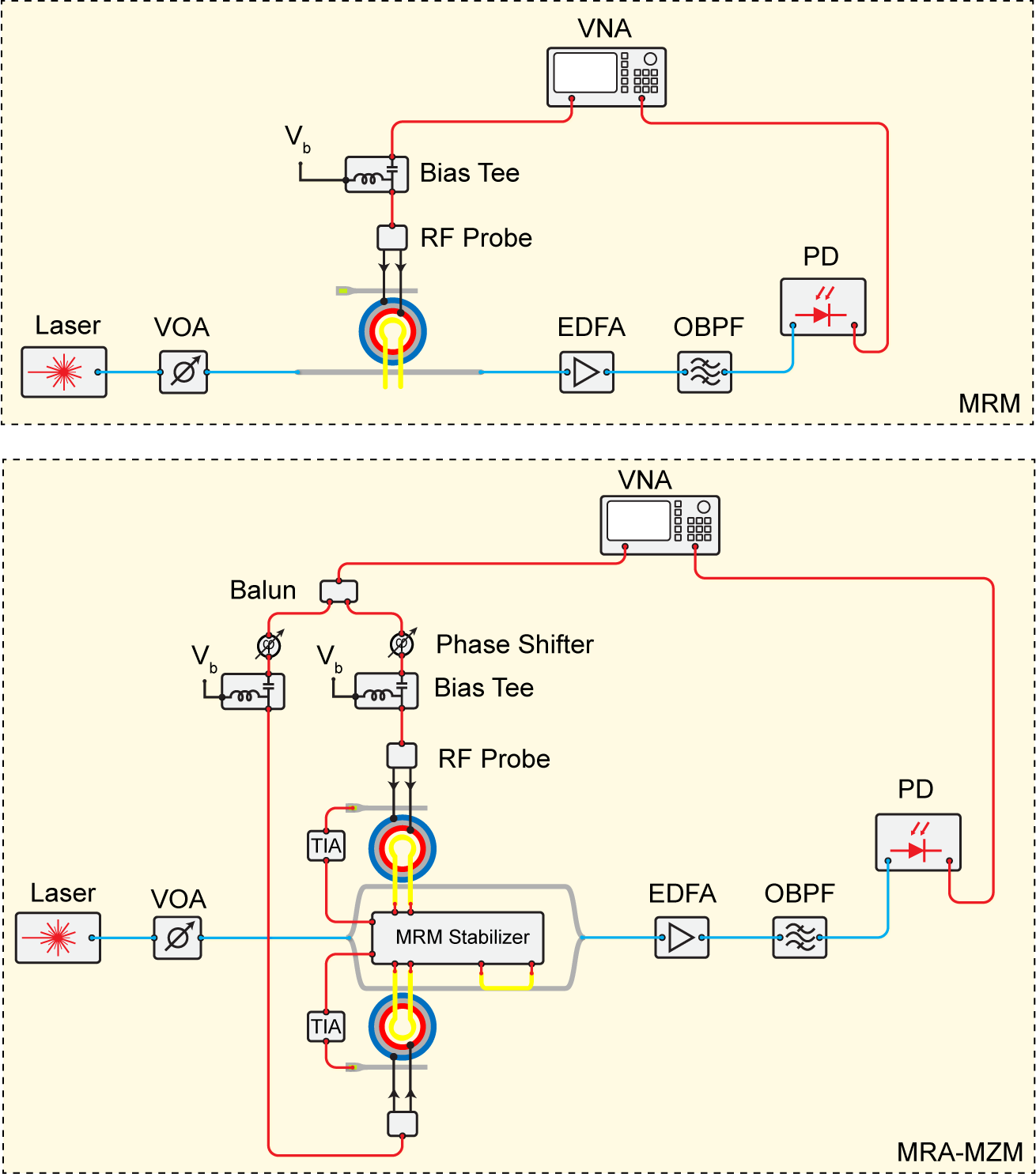}
\caption{\textbf{S-parameter characterization setups for MRM (top) and MRA-MZM (bottom).}}\label{fig:SN_eo_char}
\end{figure}

\bigskip
\newpage
\bigskip

\section{Supplementary Note: Stabilization mechanism}\label{SN:sec_stab}

On chip thermal crosstalk and ambient/chip temperature variation will impact the detuning of each ring through the experiment. In order to maintain laser-resonance alignment, we developed a closed-loop control system. For this purpose, the drop ports of the MRMs are connected to the integrated Ge-on-silicon photodetectors. Since the MRA-MZMs are operated at resonance wavelength, the drop port power needs to be maximized by shifting the spectral response of the MRM by applying heat to it through integrated TiN heaters placed on top of MRMs, aligning the resonance wavelength to the laser. The photodetector outputs are connected to off-chip logarithmic trans-impedance amplifiers to be able to cover the large dynamic range of the MRM. this monitoring signal is read by \ac{ADC} channels of the controller (NI PCIe-6361). The schematic of the closed-loop control system is presented in Fig. \ref{fig:SN_stab}.

We developed an automatic initialization and stabilization program using NI Labview to enable the control of the voltage on \ac{MRM} heaters and maintain the on-resonance operation. The developed program is designed to track the desired voltage for resonance detuning of the ring and lock on resonance by feedback information from integrated Ge-on-silicon photodetectors using a perturb and observe algorithm. The control loop was running at a 4~KHz iteration rate and was more than sufficient to capture thermal drift and maintain stability during our hours-long experiments. We apply pre-set voltage to all other \ac{MZI} heaters to meet their requirements (such as power balance, 90-degree phase shift, etc.) and then run the initialization program of MRM. Fine-tuning the \ac{MZI} heaters has negligible impact on the \ac{MRM}'s resonance because they are isolated thermally.

\begin{figure}[th]%
\centering
\includegraphics[width=0.99\textwidth]{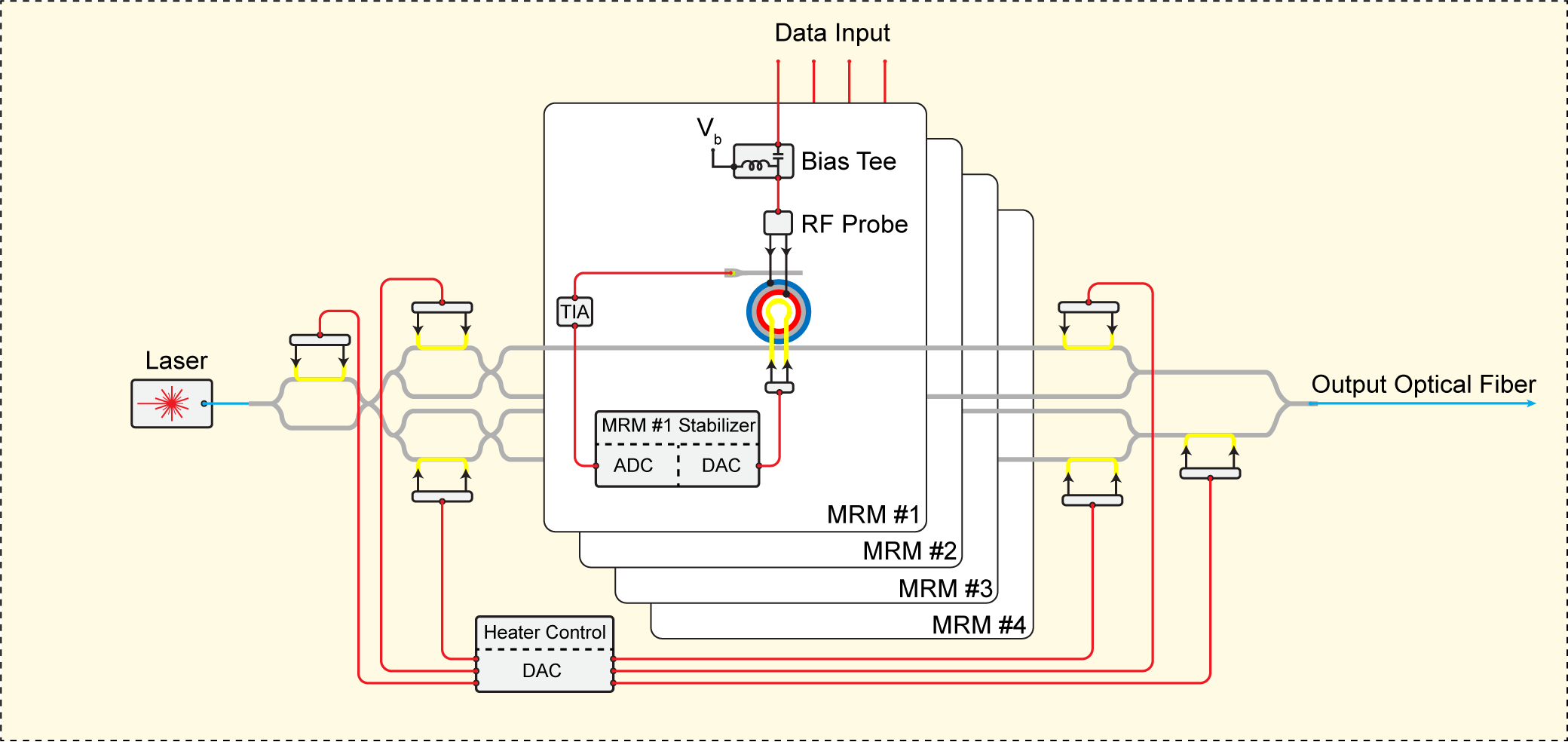}
\caption{Schematic of the developed stabilization mechanism.}\label{fig:SN_stab}
\end{figure}

\bibliographystyleS{plain}

\end{appendices}
\end{document}

\typeout{get arXiv to do 4 passes: Label(s) may have changed. Rerun}